\documentclass[pre, twocolumn, showpacs, preprintnumbers]{revtex4}

\usepackage{amssymb}
\usepackage{amsmath}
\usepackage{bm}
\usepackage{latexsym}
\usepackage{subfigmat}
\usepackage{graphicx}
\usepackage{times}
\DeclareMathAlphabet{\mathpzc}{OT1}{pzc}{m}{it}
\DeclareMathAlphabet{\mathcalligra}{T1}{calligra}{m}{n}

\begin{document}

\preprint{}
\title{Analysis of non-equilibrium fluctuations\\in a number of COVID-19 cases and deaths\\
based on time-convolutionless projection-operator method}

\author{Michio Tokuyama}
\address{Emeritus Professor, Institute of Fluid Science, Tohoku University, Sendai 980-8577, Japan}

\date{\today}

\begin{abstract}
The non-equilibrium fluctuations observed in a number of COVID-19 cases and deaths are analyzed from a statistical-dynamical point view. By investigating the data observed around the world which were collected from January 15, 2020 to April 28, 2023 at https://coronavirus.jhu.edu/, we first show that the dynamics of the fluctuations is described by a stochastic equation whose stochastic force is a multiplicative type. By employing the time-convolutionless projection-operator method in open systems previously proposed by the present author, we then transform it into a Langevin-type equation with an additive-type stochastic force together with the corresponding Fokker-Planck type equation. Thus, we explore the stochastic properties of a Langevin-type stochastic force not only analytically but also numerically from a unified point of view. Finally, we emphasize that the dynamical behavior in deaths resembles that in cases very much not only for the causal motion but also for the fluctuation.
\end{abstract}

\maketitle

\begin{table}
\caption{Fitting parameters $t_{\alpha}$, $\rho_{\alpha}$, $n_{\alpha}(t_{\alpha})$, $f_{\alpha}(t_{\alpha})$, $\gamma_{\alpha}$, and $Q_{\alpha}$ at each stage in Japan, where only the data points in the last four waves are listed for simplification.}
\begin{center}
\begin{tabular}{ccccccc}
\hline
 wave &$t_{\alpha}$ &$\rho_{\alpha}$ &$n_{\alpha}(t_{\alpha})$&$f_{\alpha}(t_{\alpha})$&$\gamma_{\alpha}\times10^3$&$Q_{\alpha}$\\
\hline
&&&-- case --&&&\\
 &525&1.0594&1.0109$\times 10^3$&7.8780$\times 10^5$&0.6156&0.2647\\
&581&0.9747&2.6026$\times 10^4$&1.2086$\times 10^6$&2.6044&0.2647\\
$[5c]$ &595&0.9148&1.8334$\times 10^4$&1.5152$\times 10^6$&0.7874&0.2647\\
&639&0.9740&3.4688$\times 10^2$&1.7178$\times 10^6$&0.6511&0.2647\\
\hline
&692&1.0773&0.8470$\times 10^2$&1.7279$\times 10^6$&1.7820&0.2647\\
&712&1.2548&3.9909$\times 10^2$&1.7304$\times 10^6$&1.7820&0.2647\\
&732&1.0550&3.8626$\times 10^4$&1.8599$\times 10^6$&2.1161&0.2647\\
$[6c]$ &749&0.9839&1.0014$\times 10^5$&2.9274$\times 10^6$&0.7204&0.2647\\
&797&1.0059&4.3964$\times 10^4$&6.1570$\times 10^6$&1.3022&0.2647\\
&824&0.9646&4.8885$\times 10^4$&7.4028$\times 10^6$&2.6044&0.2647\\
&838&1.0179&3.0972$\times 10^4$&7.9190$\times 10^6$&2.4184&0.2647\\
&853&0.9545&4.0059$\times 10^4$&8.4454$\times 10^6$&0.9958&0.2647\\
\hline
&888&1.1091&8.6064$\times 10^3$&9.1941$\times 10^6$&1.1675&0.2647\\
$[7c]$ &918&1.0014&2.0134$\times 10^5$&1.0767$\times 10^7$&1.0922&0.2647\\
&950&0.9564&2.1018$\times 10^5$&1.7398$\times 10^7$&0.8258&0.2647\\
\hline
$[8c]$ &992 &1.0197&3.2944$\times 10^4$&2.1236$\times 10^7$& 0.3680&0.2647\\
&1085&0.9474 &2.0008$\times 10^5$&2.9708$\times 10^7$&0.5209&0.2647\\
\hline
$[9c]$ &1151&1.0119&5.7928$\times 10^3$&3.3324$\times 10^7$&0.3675&0.2647\\
\hline
\hline
&&&-- death --&&&\\
&559&1.0595&8.1020&1.5151$\times 10^4$&1.3796&0.3371\\
$[5d]$ &599&0.9640&83.140&1.6300$\times 10^4$&1.3451&0.3371\\
&640&0.9385&18.234&1.8090$\times 10^4$&0.8152&0.3371\\
\hline
&707&1.1462&0.2567&1.8382$\times 10^4$&1.2228&0.3371\\
&752&1.0366&132.33&1.9207$\times 10^4$&2.9891&0.3371\\
$[6d]$ &771&0.9680&247.78&2.2880$\times 10^4$&0.9276&0.3371\\
&830&0.9972&36.187&2.9342$\times 10^4$&1.7356&0.3371\\
&862&0.9707&32.720&3.0471$\times 10^4$&1.6814&0.3371\\
\hline
&895&1.0946&6.3374&3.1281$\times 10^4$&1.4159&0.3371\\
$[7d]$ &934&1.0153&214.35&3.3460$\times 10^4$&2.3393&0.3371\\
&958&0.9578&304.43&3.9759$\times 10^4$&1.2228&0.3371\\
\hline
$[8d]$ &1003&1.0260 & 45.101 &4.5764$\times 10^4$&0.5848&0.3371\\
&1096&0.9604&487.33&6.2551$\times 10^4$&0.7079&0.3371\\
\hline
$[9d]$&1173&1.0004 & 20.720& 7.4053$\times 10^4$&0.4521&0.3371\\
\hline
\end{tabular}
\end{center}
\label{table-1}
\end{table}
\begin{figure}
\begin{center}
\includegraphics[width=10.0cm]{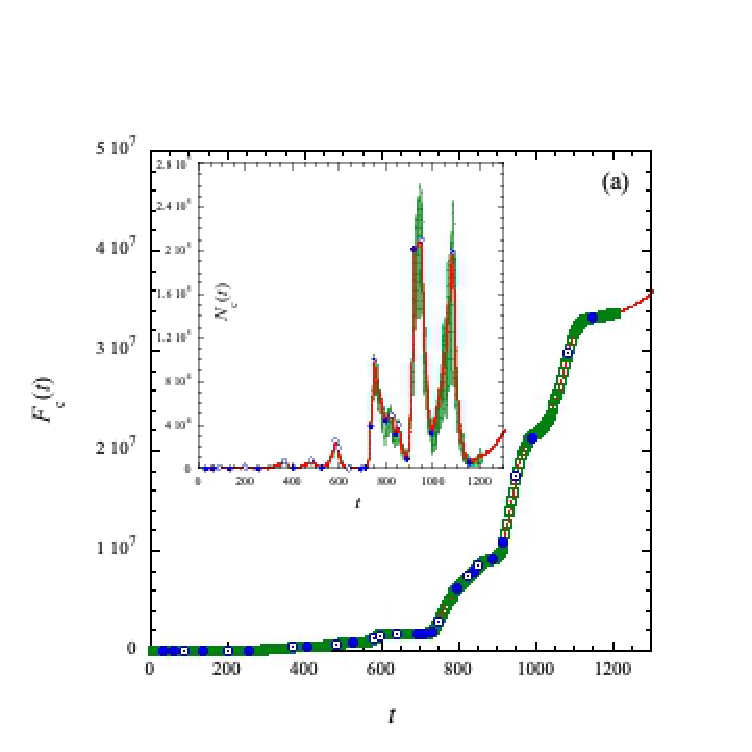}
\includegraphics[width=10.0cm]{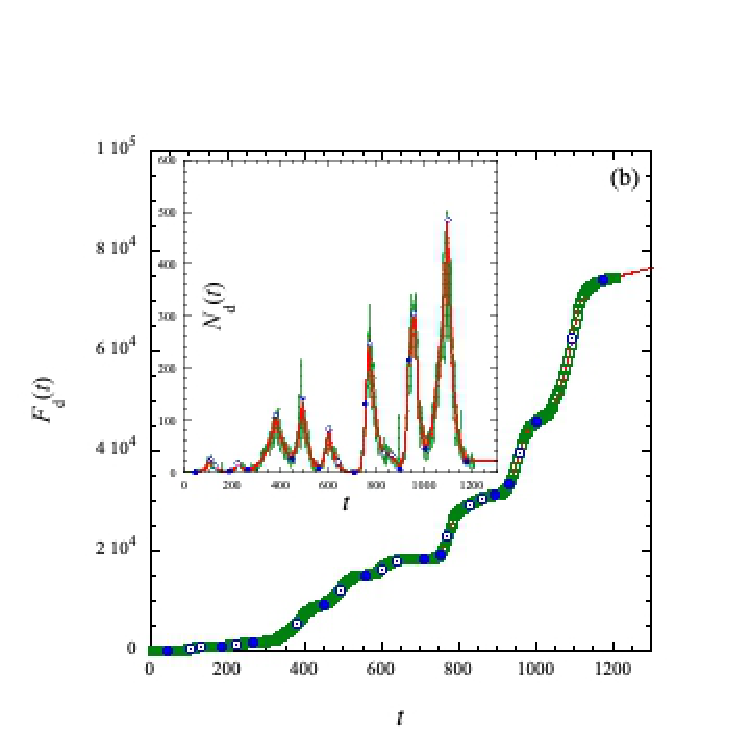}
\end{center}
\caption{(Color online) A plot of the case and death counts versus time $t$ in Japan. (a) a cumulative number $F_c(t)$ for cases and (b) a cumulative number $F_d(t)$ for deaths, where the open squares indicate the observed data in Japan and the solid line the averaged cumulative number $f_{\alpha}(t)$. The symbols ($\bullet$) indicate the crossover times with $\lambda_{\alpha}>0$ and ($\odot$) with $\lambda_{\alpha}<0$, whose values are listed in Table \ref{table-1}. For comparison, the fluctuating daily number $N_{\alpha}(t)$ is also given in the inset, where the dotted line indicates the observed data and the solid line the averaged daily number $n_{\alpha}(t)$.}
\label{jp1}
\end{figure}
\begin{figure}
\begin{center}
\includegraphics[width=10.0cm]{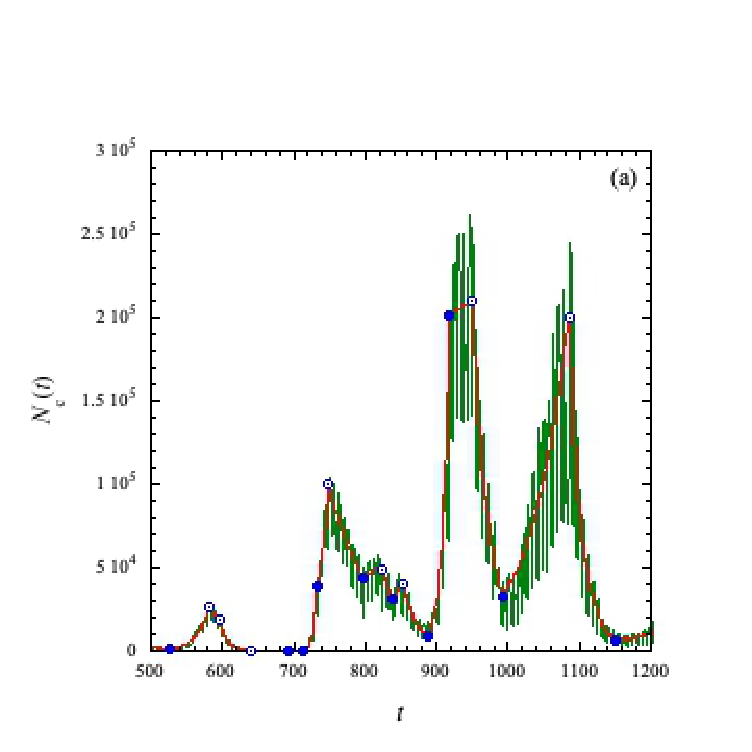}
\includegraphics[width=10.0cm]{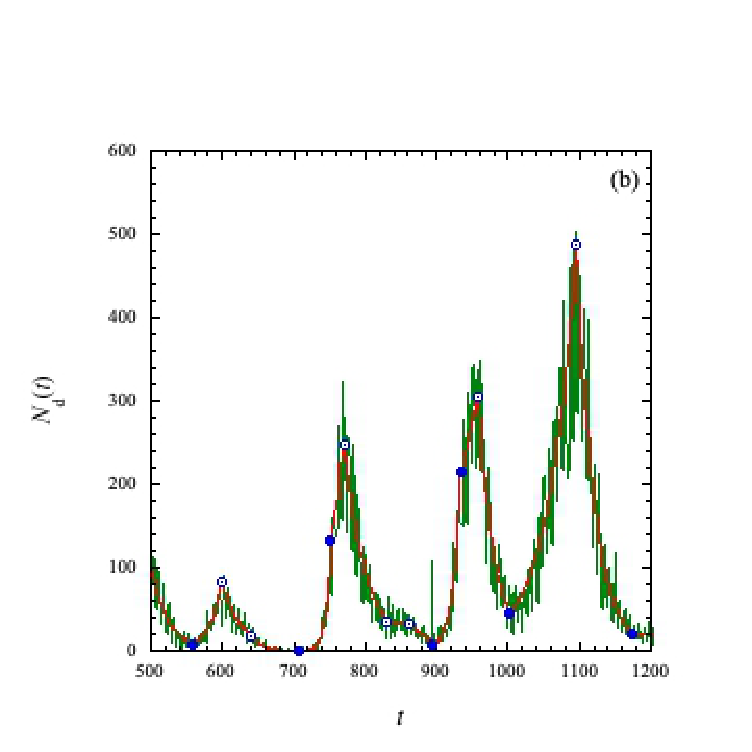}
\end{center}
\caption{(Color online) A plot of the case and death daily counts observed in Japan versus time $t$ around the last three peaks for $500\leq t \leq 1200$, where (a) $N_c(t)$ and (b) $N_d(t)$. The fluctuating line indicates the observed daily counts given by $N_{\alpha}(t)$. The details are the same as in Fig. \ref{jp1}. }
\label{jp2}
\end{figure}
\section{Introduction}
A mathematical model (the so-called SIR model) to describe an epidemic outbreak of an infectious disease by using a differential equation has been first proposed by Kermack and McKendrick \cite{KM27,KM32,DHB13}. In order to describe the average number $n_c(t)$ of COVID-19 daily case counts at time $t$, therefore, we employ an exponential function given by $n_c(t)=n_c(t_c)\rho_c^{(t-t_c)}$ based on such a simple model, where $\rho_c$ is an infection rate and $t_c$ is a crossover time when the infection mechanism with $\rho_c$ starts to work. By analyzing many different data observed in various countries, we then show that the same function as that for $n_c(t)$ also holds for the average number $n_d(t)$ of COVOD-19 daily death counts, except that the infection rate $\rho_c$ is now replaced by the fatality rate $\rho_d$. Thus, $n_{\alpha}(t)$ is assumed to obey an exponential function given by $n_{\alpha}(t)=\rho_{\alpha}^{(t-t_{\alpha})}n_{\alpha}(t_{\alpha})=e^{\lambda_{\alpha} (t-t_{\alpha})}n_{\alpha}(t_{\alpha})$, where $\lambda_{\alpha}=\ln(\rho_{\alpha})$, and $\alpha=c$ stands for cases and $\alpha=d$ for deaths. Here $t_{\alpha}$ indicates a crossover time when the infection mechanism with $\rho_{\alpha}$ starts to work. By using $n_{\alpha}(t)$, one can then formally obtain the average number $f_{\alpha}(t)$ of cumulative case and death counts as $f_{\alpha}(t)=[n_{\alpha}(t)-n_{\alpha}(t_{\alpha})]/\lambda_{\alpha}+f_{\alpha}(t_{\alpha})$. When $\rho_{\alpha}>1$, both functions grow with the growth rate $\lambda_{\alpha} (>0)$ as $t$ increase. On the other hand, when $\rho_{\alpha}<1$, they decrease with the decay rate $\lambda_{\alpha} (<0)$. First, we check how both mean functions work well to describe the COVID-19 cases and deaths appeared in different countries from a unified point of view. Then, we investigate the dynamics of non-equilibrium fluctuations $\delta n_{\alpha}(t)$ around $n_{\alpha}(t)$ and also $\delta f_{\alpha}(t)$ around $f_{\alpha}(t)$. Then, we explore their stochastic properties from a statistical-dynamical point of view. Thus, we confirm that the fluctuating daily number $N_{\alpha}(t)(=n_{\alpha}(t)+\delta n_{\alpha}(t))$ must be described by a multiplicative-type stochastic equation. By employing the time-convolutionless projection-operator method \cite{toku75,toku76}, we then show that it can be transformed into a Langevin-type equation with an additive-type stochastic force together with the corresponding Fokker-Planck type equation for $P(n_{\alpha},t)$, which is a probability distribution function for $N_{\alpha}(t)$ to have a value $n_{\alpha}$ at time $t$. In order to confirm the validity of the proposed stochastic equation, we have analyzed the various data observed in different countries from a unified point of view. Thus, we show how such a Langevin-type equation can describe not only the causal motion of number in COVID-19 cases and deaths in different countries but also the dynamics of the non-equilibrium fluctuations around the causal motion. We finally emphasize that the dynamical behavior in deaths is very similar to that in cases not only for the causal motion but also for the fluctuation. In the present paper, we only report the main results obtained by using the data observed in Japan on account of the space. In summary, however, we just compare the data for the fluctuating cumulative number $F_{\alpha}(t)(=f_{\alpha}(t)+\delta f_{\alpha}(t))$ observed in various countries with the analytic results for $f_{\alpha}(t)$ and show how analytic results can describe the fluctuating numbers well in any countries. Thus, the fluctuating death rate given by $F_d(t)/F_c(t)$ in each country is also shown to be well described by the analytic one given by $f_d(t)/f_c(t)$.

\section{Starting equations}
In this section, we first discuss the deterministic equations to describe the dynamics of number in COVID-19 cases and deaths in different countries. Let $\rho_c$ be an infection rate, where one infected person can infect the $\rho_c$ persons at one day. Let $N_c(t)$ and $F_c(t)$ also be a fluctuating daily number and a fluctuating cumulative number, respectively, which are observable quantities at time $t$ in various countries. Then, one can write their averaged numbers as $n_c(t)=\langle N_c(t)\rangle$ and $f_c(t)=\langle F_c(t)\rangle$, where the brackets denote the average over an appropriate initial ensemble. After $t$ days, therefore, the mean number of infected persons is given by $n_c(t)\propto\rho_c^t$. The many data observed in different countries also suggest that there exists an obvious correlation between the dynamics of case counts and that of death counts since their dynamics show a similar behavior in time. Introducing the fatality rate by $\rho_d$, one may also find the daily number of dead persons after $t$ days as $n_d\propto\rho_d^t$. The observed data for daily count are always fluctuating in time. Those fluctuations are not avoidable because the infected persons are always surrounded by many other persons and also because the social conditions are also changing in time. Such a situation is similar to that of a Brownian motion in a heat bath. We first discuss the analytic function of $t$ for $n_{\alpha}(t)$, where $\alpha=c$ stands for cases and $\alpha=d$ for deaths. Let $t_{\alpha}$ denote the crossover time when the mechanism with $\rho_{\alpha}$ starts to work. Here we note that $t_{\alpha}$ is not necessarily a first day ($t=1$) when the first COVID-19 cases (or deaths) appear. The mean cumulative number $f_{\alpha}(t)$ is given as
\begin{equation}
f_{\alpha}(t)=\int_{t_{\alpha}}^t n_{\alpha}(s)ds+f_{\alpha}(t_{\alpha}).\label{NF}
\end{equation}
The average number for daily $\alpha$ count is then assumed to be described by
\begin{equation}
n_{\alpha}(t)=\rho_{\alpha}^{(t-t_{\alpha})}n_{\alpha}(t_{\alpha})
=e^{\lambda_{\alpha}(t-t_{\alpha})}n_{\alpha}(t_{\alpha}), \label{dnum}
\end{equation}
where $\lambda_{\alpha}=\ln(\rho_{\alpha})$. Use of Eqs. (\ref{NF}) and (\ref{dnum}) leads to
\begin{equation}
f_{\alpha}(t)=[e^{\lambda_{\alpha}(t-t_{\alpha})}-1]n_{\alpha}(t_{\alpha})/\lambda_{\alpha}+f_{\alpha}(t_{\alpha}). \label{cnum}
\end{equation}
Hence the causal motions for $n_{\alpha}(t)$ and $f_{\alpha}(t)$ are turned out to obey the following deterministic equations:
\begin{eqnarray}
\frac{d}{dt}n_{\alpha}(t)&=&\lambda_{\alpha}n_{\alpha}(t), \label{dnumt}\\
\frac{d}{dt} f_{\alpha}(t)&=&n_{\alpha}(t),\label{cnumt}
\end{eqnarray}
respectively. Here the unknown parameters $\rho_{\alpha}$, $t_{\alpha}$, $n_{\alpha}(t_{\alpha})$, and $f_{\alpha}(t_{\alpha})$ are determined consistently by fitting $f_{\alpha}(t)$ with observed data in each community. We note here that the rate $\rho_{\alpha}$ is sensitive to the social conditions and $\lambda_{\alpha}=0$ corresponds to a crossover point from a growth process with $\lambda_{\alpha}>0$ (or $\rho_{\alpha}>1)$ to a decay process with $\lambda_{\alpha}<0$ (or $0<\rho_{\alpha}<1)$ and vice versa. This will be discussed later.

As a typical example, we first show how Eqs. (\ref{dnum}) and (\ref{cnum}) can describe the data observed in Japan well, which were collected from the first day January 16, 2020 ($t=1$) to May 8, 2023 ($t=1209$). 
In Fig. \ref{jp1}, the observed data for case and death counts are shown together with Eqs. (\ref{dnum}) and (\ref{cnum}). They are shown to be well described by those analytic functions for all times, where the fitting values of the parameters $\rho_{\alpha}$, $n_{\alpha}(t_{\alpha})$, and $f_{\alpha}(t_{\alpha})$ are listed in Table \ref{table-1}. Up to May 8, 2023, there are mainly 8 waves for both case and death in Japan. In a typical wave, $n_{\alpha}(t)$ starts to increase from the minimum value $n_{\alpha}(t_{\alpha})$ to the maximum value $n_{\alpha}(t_{\alpha}')$ and then to decrease to the minimum value $n_{\alpha}(t_{\alpha}")$. Here we also call such a region a stage during which $\rho_{\alpha}$ is constant. Let $t_{\alpha \ell}^{(i)}$ be a crossover time at the stage $i$ in the wave $\ell$. For example, in the first wave one finds $t_{c1}^{(1)}=30$ and $t_{d1}^{(1)}=43$. In the 5th wave [5c] $n_c(t)$ starts to grow at $t=t_{c5}^{(1)}(=525)$ and then to decrease at $t=t_{c5}^{(2)}(=581)$ up to $t=t_{c6}^{(1)}-1(=691)$, while in the 5th stage [5d] $n_d(t)$ starts to grow at $t=t_{d5}^{(1)}(=559)$ and then to decrease at $t=t_{d5}^{(2)}(=599)$ up to $t=t_{d6}^{(1)}-1(=706)$. For example, there are 3 stages in [7$\alpha$], while 8 stages in [6c] and 5 stages in [6d]. Thus, it turns out that the dynamics of death counts is very similar to that of case counts and the crossover time $t_d$ is always behind the corresponding time $t_c$ (see Table \ref{table-1}). This is very reasonable because people will die after they will be infected with COVID-19. The accelerating and decelerating processes for case and death counts are thus repeated, forming several waves, until the humankind will prevail against the corona viruses. Those situations are shown to hold for any data observed not only in different countries but also in different communities, such as states, prefectures, cities, towns, etc. 
\begin{figure}
\begin{center}
\includegraphics[width=11.0cm]{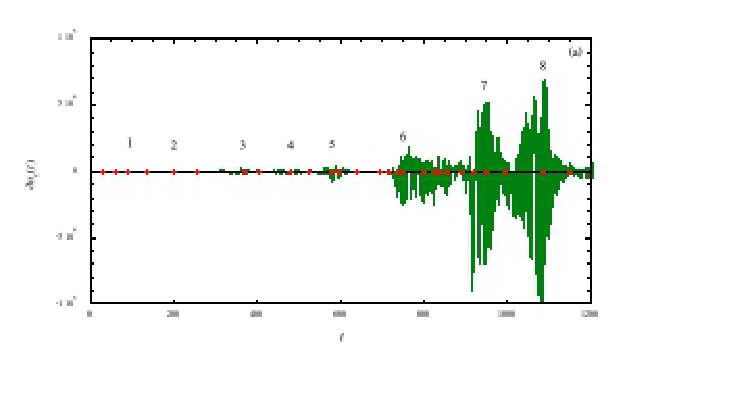}
\includegraphics[width=11.0cm]{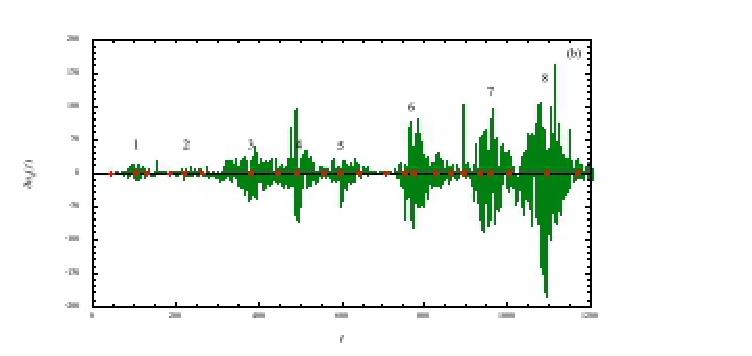}
\end{center}
\caption{(Color online) A plot of the fluctuations $\delta n_{\alpha}(t)$ observed in Japan versus time $t$, where (a) case and (b) death. The solid line indicates $\delta n_{\alpha}(t)$. The number $i$ indicates the crossover time of the peak in the wave [$i_{\alpha}$]. The details are the same as in Fig. \ref{jp1}.}
\label{jp3}
\end{figure}
\begin{figure}
\begin{center}
\includegraphics[width=11.0cm]{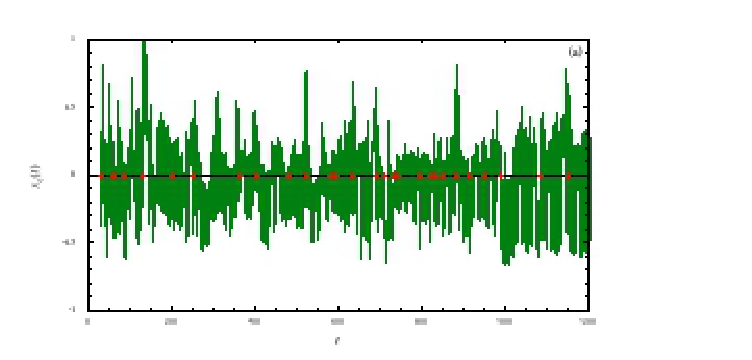}
\includegraphics[width=11.0cm]{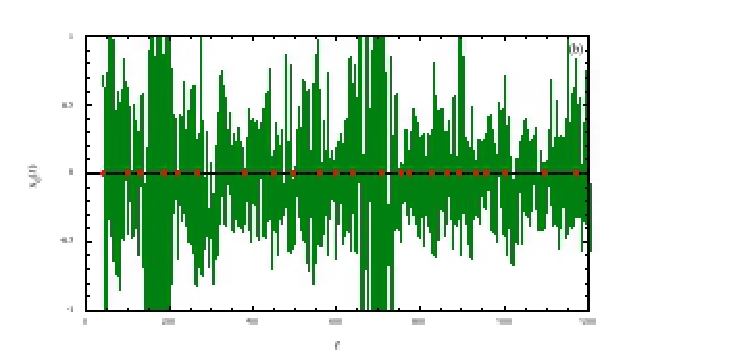}
\end{center}
\caption{(Color online) A plot of the fluctuations $x_{\alpha}(t)$ observed in Japan versus time $t$, where (a) $x_c(t)$ and (b) $x_d(t)$. The solid lines indicate $x_{\alpha}(t)$. The details are the same as in Fig. \ref{jp1}.}
\label{jp4}
\end{figure}
\begin{figure}
\begin{center}
\includegraphics[width=11.0cm]{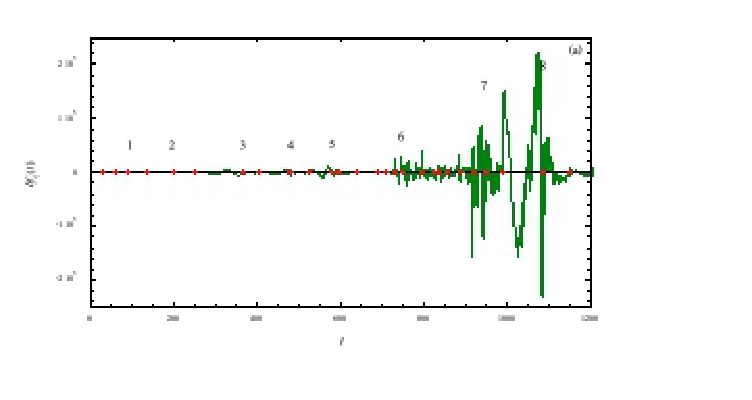}
\includegraphics[width=11.0cm]{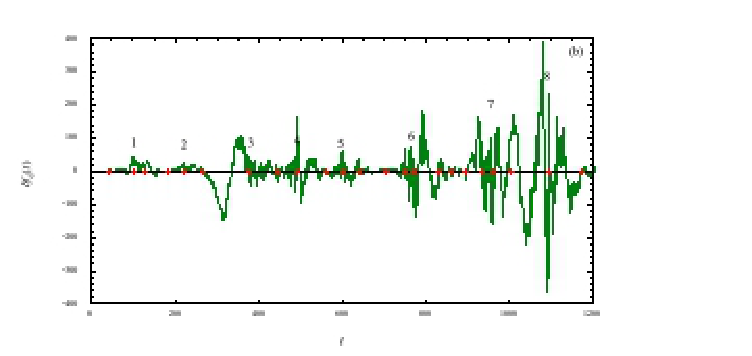}
\end{center}
\caption{(Color online) A plot of the fluctuations $\delta f_{\alpha}(t)$ observed in Japan versus time $t$, where (a) $\delta f_c(t)$ and (b) $\delta f_d(t)$. The solid lines indicate $\delta f_{\alpha}(t)$. The details are the same as in Figs. \ref{jp1} and \ref{jp3}.}
\label{jp5}
\end{figure}
\begin{figure}
\begin{center}
\includegraphics[width=11.0cm]{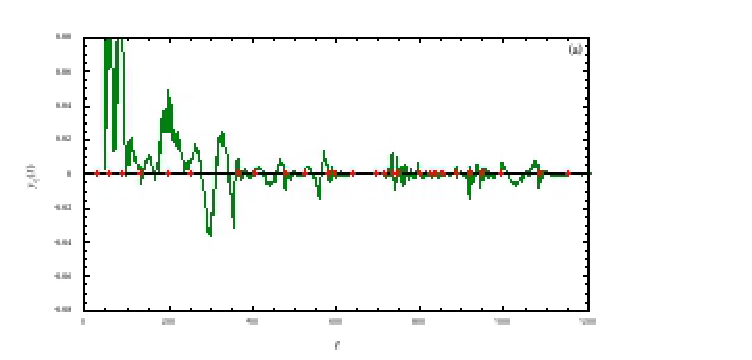}
\includegraphics[width=11.0cm]{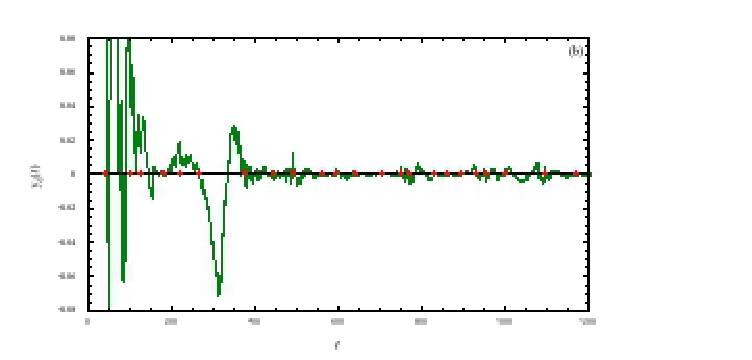}
\end{center}
\caption{(Color online) A plot of the fluctuations $y_{\alpha}(t)$ observed in Japan versus time $t$, where (c) $y_c(t)$ and (d) $y_d(t)$. The solid lines indicate $y_{\alpha}(t)$. The details are the same as in Fig. \ref{jp1}.}
\label{jp6}
\end{figure}
\begin{figure}
\begin{center}
\includegraphics[width=10.5cm]{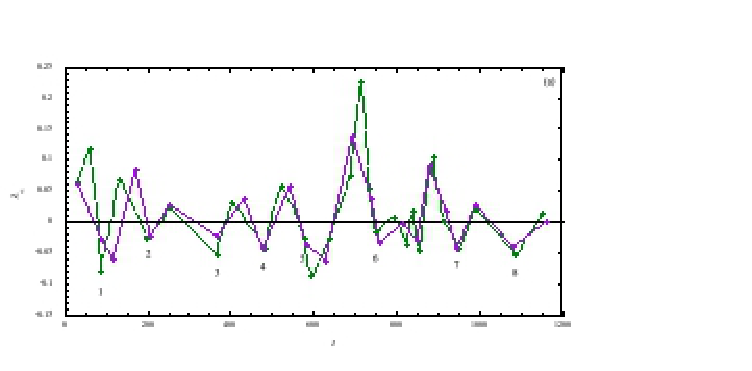}
\includegraphics[width=10.5cm]{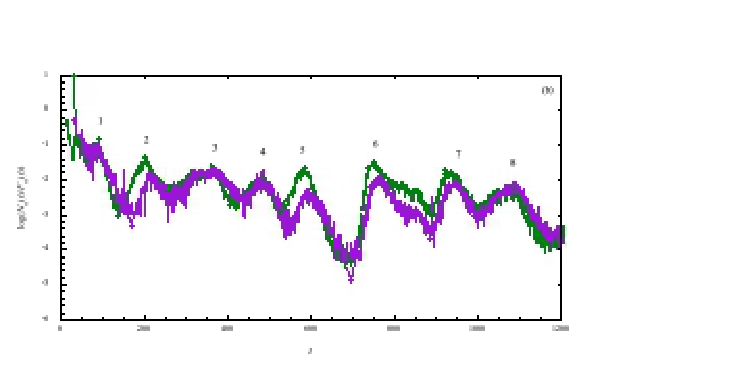}
\includegraphics[width=10.5cm]{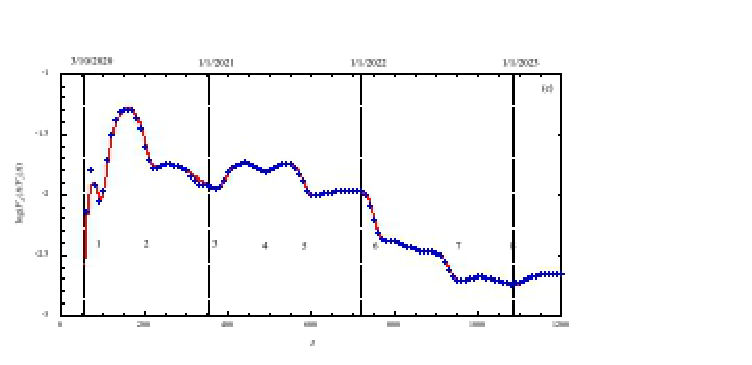}
\end{center}
\caption{(Color online) A log plot of (a) the infection rate $\rho_{\alpha}$ (or $\lambda_{\alpha}$), (b) the ratio $N_{\alpha}(t)/F_{\alpha}(t)$, and (c) the death rate $F_d(t)/F_c(t)$ versus time $t$. In (a) the solid line indicates $\lambda_d$ and the dashed line $\lambda_c$. In (b) the fluctuating solid line indicates the ratio $N_d(t)/F_d(t)$ and the fluctuating dashed line $N_c(t)/F_c(t)$, where the solid line indicates the ratio $n_d(t)/f_d(t)$ and the dashed line the ration $n_c(t)/f_c(t)$. In (c) the symbols ($\bullet$) indicate the ratio $F_d(t)/F_c(t)$ and the solid line the ratio $f_d(t)/f_c(t)$, where the vertical dashed lines the calendar days. The starting time $t_{d1}^{(1)}(=43)$ is adjusted to coincide with $t_{c1}^{(1)}(=30)$. The details are the same as in Figs. \ref{jp1} and \ref{jp3}.}
\label{jp7}
\end{figure}

There exist the non-equilibrium fluctuations $\delta n_{\alpha}(t)$ and $\delta f_{\alpha}(t)$ around their mean values $n_{\alpha}(t)$ and $f_{\alpha}(t)$, respectively. In fact, the fluctuations $\delta n_{\alpha}(t)$ are clearly seen in Figs. \ref{jp1} and \ref{jp2}, while the fluctuations $\delta f_{\alpha}(t)$ are not clearly found because their magnitudes are very small as will be discussed below. In Fig. \ref{jp3}, the numerical values of $\delta n_{\alpha}(t)$ are shown for $1\leq t \leq 1200$. They are the non-equilibrium fluctuations, which  must be caused under the complicated social conditions. Hence their magnitudes are enhanced around the peaks of $N_{\alpha}(t)$. Here it is convenient to introduce their relative magnitudes by $x_{\alpha}(t)=\delta n_{\alpha}(t)/n_{\alpha}(t)$ where $\langle x_{\alpha}(t)\rangle=0$. In Fig. \ref{jp4}, those are also plotted versus time for $1\leq t \leq 1200$. They are shown to fluctuate from -1.0 to 1.0, except errors.  Here we note that $x_{\alpha}(t)$ are not random in time since the system is not in equilibrium. We also introduce the relative magnitudes of the fluctuations $\delta f_{\alpha}(t)$ by $y_{\alpha}(t)=\delta f_{\alpha}(t)/f_{\alpha}(t)$, where $\langle y_{\alpha}(t)\rangle=0$. In Figs. \ref{jp5} and \ref{jp6}, $\delta f_{\alpha}(t)$ and $y_{\alpha}(t)$ are plotted versus time for $1\leq t \leq 1200$, respectively. The magnitudes of $\delta f_{\alpha}(t)$ are also shown to be enhanced around the peaks of $N_{\alpha}(t)$. The non-equilibrium fluctuations $y_{\alpha}(t)$ are shown to fluctuate mainly from -0.08 to 0.08, except errors. Hence we have $|y_{\alpha}(t)|\ll|x_{\alpha}(t)|\leq 1$. This is the reason why $\delta f_{\alpha}(t)$ is not seen clearly in Fig. \ref{jp1} as compared to $\delta n_{\alpha}(t)$. The figures \ref{jp3}-\ref{jp6} ensure that the dynamical behaviors for death counts are very similar to that of case counts. In order to see this clearly, in Figs. \ref{jp7} (a), (b), and (c)  we further compare the dynamics in death with that in case by using the infection rate $\rho_{\alpha}$, the ratio $N_{\alpha}(t)/F_{\alpha}(t)$, and the death rate $F_d(t)/F_c(t)$, respectively. Then, those figures further confirm that their dynamical behaviors are very similar to each other. Although we have used the data observed in Japan in order to check such a similarity, it is also true for the data observed in different countries. Hence $N_d(t)$ (or $F_d(t)$) is expected to be described by the same equation as that for $N_c(t)$ (or $F_c(t)$). In the next section, therefore, we first discuss the stochastic equation for $N_{\alpha}(t)$ to describe the non-equilibrium fluctuations for case and death and then investigate their stochastic properties together with those for $F_{\alpha}(t)$ from a unified point of view. 

\section{Stochastic equations in non-equilibrium systems}
In this section, we first propose a multiplicative-type stochastic equation for $N_{\alpha}(t)$. By employing the time-convolutionless projection operator method \cite{toku75,toku76}, we then transform it into a Langevin-type equation together with a corresponding Fokker-Planck type equation for $P(n_{\alpha},t)$, which is a probability distribution function for $N_{\alpha}(t)$ to have a value $n_{\alpha}$ at time $t$. Thus, we investigate the stochastic properties of the non-equilibrium fluctuations not only analytically but also numerically from a unified point of view. 
 
\subsection{A starting stochastic equation}
In order to discuss the dynamics of the fluctuations, one may start from the stochastic equation given by
\begin{equation}
\frac{d}{dt}N_{\alpha}(t)=\lambda_{\alpha 0} N_{\alpha}(t)+\Omega_{\alpha}(t), \label{Meq}
\end{equation}
where $\Omega_{\alpha}(t)$ is a non-equilibrium stochastic force and $\lambda_{\alpha 0}$ a bare infection rate without noise. Similarly to Eq. (\ref{cnumt}), the fluctuating cumulative number $F_{\alpha}(t)$ is also described by
\begin{equation}
\frac{d}{dt}F_{\alpha}(t)=N_{\alpha}(t). \label{dFdt}
\end{equation}
We next explore the stochastic force $\Omega_{\alpha}(t)$. If one takes the average of Eqs. (\ref{Meq}) and (\ref{dFdt}), they should reduce to Eqs. (\ref{dnumt}) and (\ref{cnumt}), respectively. Hence we obtain
\begin{equation}
\langle \Omega_{\alpha}(t)\rangle=\gamma_{\alpha}n_{\alpha}(t), \label{avome}
\end{equation}
where $\gamma_{\alpha}(=\lambda_{\alpha}-\lambda_{\alpha 0})$ is a positive constant to be determined on each stage. Thus, it turns out that the stochastic force $\Omega_{\alpha}(t)$ seems to enhance the bare rate $\lambda_{\alpha 0}$. As discussed in the previous papers \cite{toku80,toku81}, this suggests that $\Omega_{\alpha}(t)$ could be a multiplicative-type stochastic force. In fact, as is shown in Appendix A, one can write $\Omega_{\alpha}(t)$ as
\begin{figure}
\begin{center}
\includegraphics[width=11.0cm]{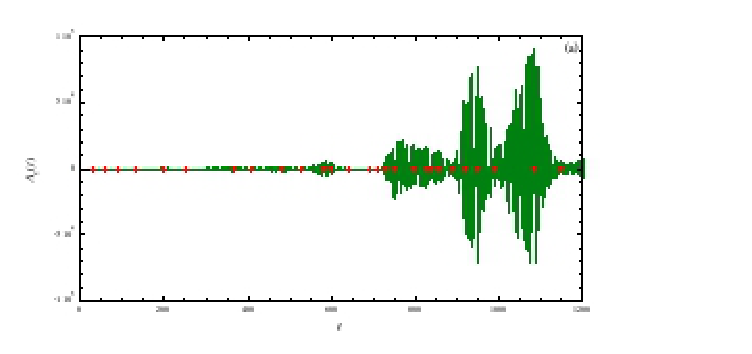}
\includegraphics[width=11.0cm]{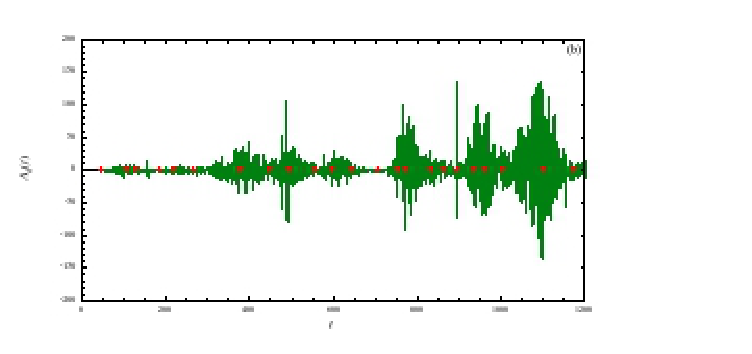}
\includegraphics[width=11.0cm]{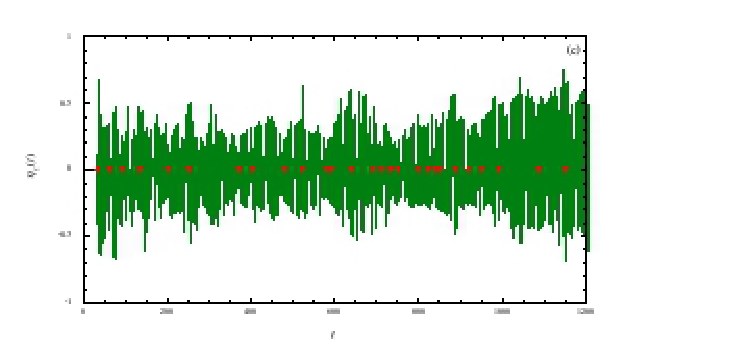}
\includegraphics[width=11.0cm]{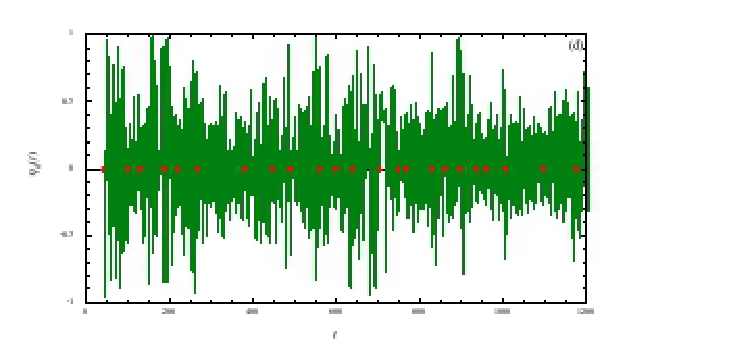}
\end{center}
\caption{(Color online) A plot of the stochastic forces $\Lambda_{\alpha}(t)$ and $\eta_{\alpha}(t)$ versus time $t$, where (a) $\Lambda_c(t)$, (b) $\Lambda_d(t)$, (c) $\eta_c$, and (d) $\eta_d$. The solid lines indicate the numerical results obtained by using the data observed in Japan. The details are the same as in Fig.\ref{jp1}.}
\label{jp8}
\end{figure}
\begin{figure}
\begin{center}
\includegraphics[width=10.5cm]{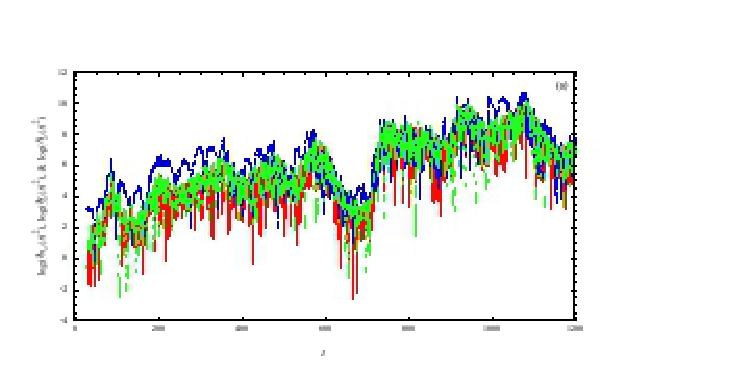}
\includegraphics[width=10.5cm]{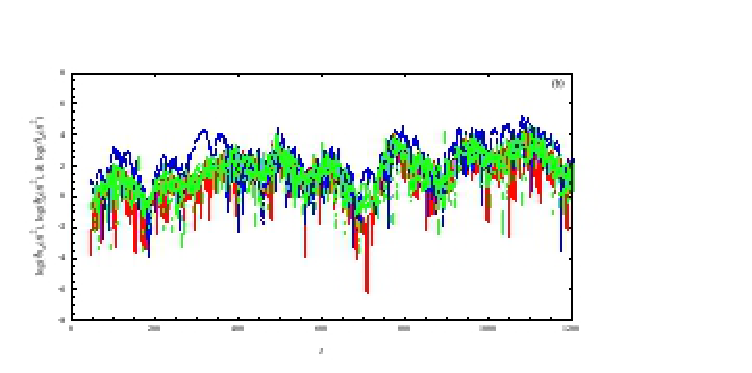}
\includegraphics[width=10.5cm]{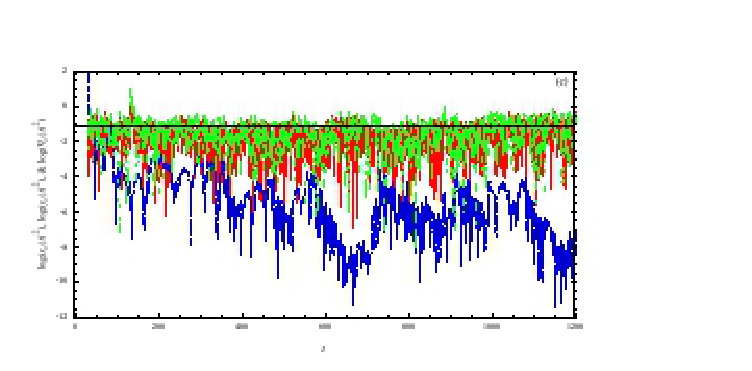}
\includegraphics[width=10.5cm]{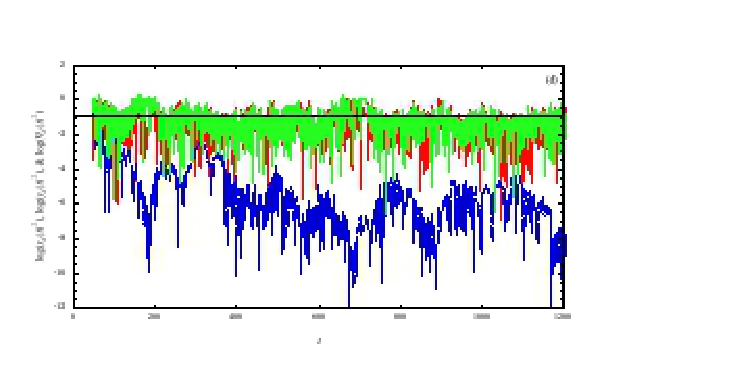}
\end{center}
\caption{(Color online) A log plot of the square of the observed data versus time $t$. (a) $\delta n_c(t)^2$, $\delta f_c(t)^2$, and $\Lambda_c(t)^2$, (b) $\delta n_d(t)^2$, $\delta f_d(t)^2$, and $\Lambda_d(t)^2$, (c) $x_c(t)^2$, $y_c(t)^2$, and $\eta_c(t)^2$, and (d) $x_d(t)^2$, $y_d(t)^2$, and $\eta_d(t)^2$. The fluctuating red solid lines indicate the observed data for $\Lambda_{\alpha}(t)^2$, the fluctuating green dotted lines for $\delta n_{\alpha}(t)^2$, and the fluctuating blue dashed lines for $\delta f_{\alpha}(t)^2$. The bold horizontal lines indicate the mean square $Q_{\alpha}^2$ of $x_{\alpha}(t)^2$, where $Q_c\simeq 0.2647$ and $Q_d\simeq 0.3371$.}
\label{jp9}
\end{figure}
\begin{figure}
\begin{center}
\includegraphics[width=9.0cm]{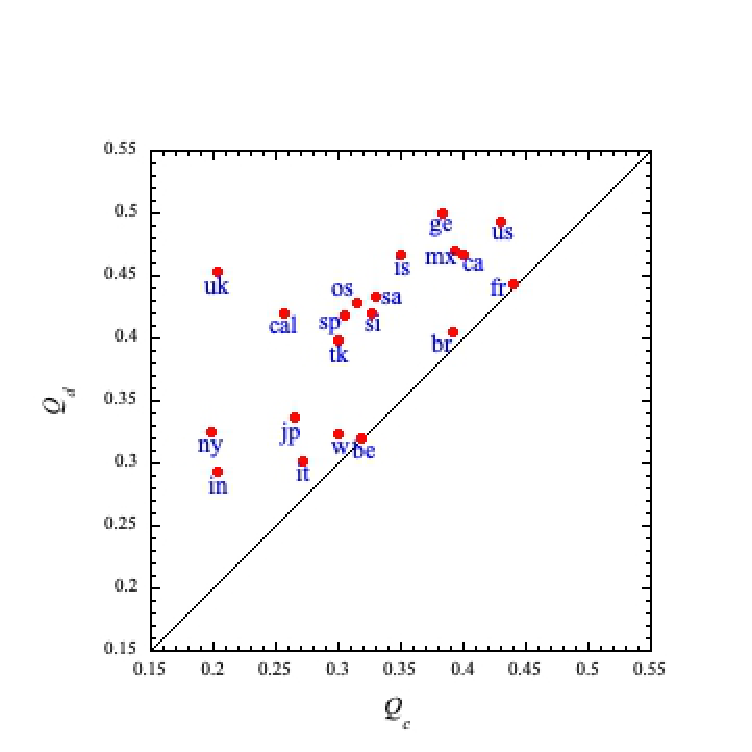}
\end{center}
\caption{(Color online) A plot of $Q_d$ versus $Q_c$ for different communities. The symbols ($\bullet$) indicate the value of $Q_{\alpha}$ at different communities. The abbreviations indicate the communities as be (Belgium), br (Brazil), ca (Canada), cal (California state), fr (France), ge (Germany), in (India), is (Israel), it (Italy), jp (Japan),  mx (Mexico), ny (New York state), os (Osaka pref.), sa (South Africa), si (Singapore), sp ( Spain), tk (Tokyo pref.), uk (United Kingdom), us (USA), and w (World). The solid line indicates the straight line given by $Q_d=Q_c$.   }
\label{jp10}
\end{figure}
\begin{figure}
\begin{center}
\includegraphics[width=11.5cm]{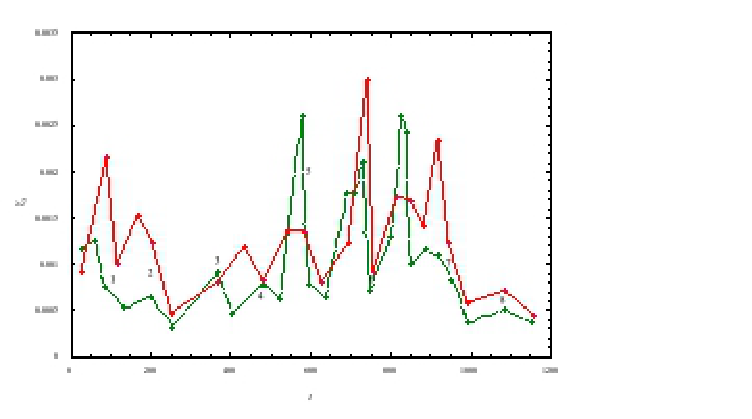}
\end{center}
\caption{(Color online) A plot of $\gamma_{\alpha}$ versus time $t$. The solid line indicates $\gamma_d$ and the dashed line $\gamma_c$. The number $i$ indicates the crossover time of the peak in the wave [$\ell_{\alpha}$]. The details are the same as in Fig. \ref{jp7}.}
\label{jp11}
\end{figure}
\begin{equation}
\Omega_{\alpha}(t)=\xi_{\alpha}(t)N_{\alpha}(t), \label{Ome}
\end{equation}
where $\xi_{\alpha}(t)$ is a stochastic force per a single person. In order to explore the stochastic properties of $\xi_{\alpha}(t)$, we first check the function $\Lambda_{\alpha}(t)$ given by
\begin{equation}
\Lambda_{\alpha}(t)=\Omega_{\alpha}(t)-\gamma_{\alpha}N_{\alpha}(t)= \frac{dN_{\alpha}(t)}{dt}-\lambda_{\alpha}N_{\alpha}(t), \label{nome}
\end{equation}
where $\langle \Lambda_{\alpha}(t)\rangle=0$. In the next subsection, this is shown to be an additive-type stochastic force $R_{\alpha}(t)$ in a new Langevin-type equation. The numerical values of $\Lambda_{\alpha}(t)$ could be obtained by using the observed data. In fact, they can found numerically from Eq. (\ref{nome}) by using a fourth-order center difference method to approximate a derivative of $N_{\alpha}(t)$. The numerical results for $\Lambda_{\alpha}(t)$ obtained by using the data observed in Japan are then plotted versus time in Figs. \ref{jp8} (a) and (b). The time behaviors of $\Lambda_{\alpha}(t)$ are very similar to those of $\delta n_{\alpha}(t)$ shown in Fig. \ref{jp3}, where their magnitudes are also enhanced around the peaks of $N_{\alpha}(t)$. In order to see the relation between $\Lambda_{\alpha}(t)$ and $x_{\alpha}(t)$, it is convenient to introduce a single function $\eta_{\alpha}(t)$ by $\eta_{\alpha}(t)=\Lambda_{\alpha}(t)/n_{\alpha}(t)$. In Figs. \ref{jp8} (c) and (d), the numerical results for $\eta_{\alpha}(t)$ are also shown. Their time behaviors seem to be rather random compared to those of $x_{\alpha}(t)$ (see Fig. \ref{jp4}). This will be discussed later. 

Finally, we discuss the fluctuations $\delta f_{\alpha}(t)$. From Eq. (\ref{dFdt}), one can write $\delta f_{\alpha}(t)$ in terms of $\delta n_{\alpha}(t)$ as
\begin{equation}
\delta f_{\alpha}(t)=\delta f_{\alpha}(t_{\alpha})+\int_{t_{\alpha}}^t \delta n_{\alpha}(s)ds. \label{yt}
\end{equation}
This also suggests that the dynamical behavior of $\delta f_{\alpha}(t)$ must be similar to that of $\delta n_{\alpha}(t)$. In order to see this clearly, in Figs. \ref{jp9} (a) and (b) we directly compare the fluctuations $\delta n_{\alpha}(t)^2$, $\delta f_{\alpha}(t)^2$, and $\Lambda_{\alpha}(t)^2$ with each other, which are obtained by the observed data. Their dynamical behaviors are shown to be similar to each other. As is discussed later, such a similarity results from the fact that both the fluctuations $\delta n_{\alpha}(t)$ and $\delta f_{\alpha}(t)$ can be written in terms of the additive-type stochastic forces $R_{\alpha}(t)$ (or $\Lambda_{\alpha}(t)$). In Figs. \ref{jp9} (c) and (d), we also compare the fluctuations $x_{\alpha}(t)^2$, $y_{\alpha}(t)^2$, and $\eta_{\alpha}(t)^2$ with each other. Their dynamical behaviors are also shown to be similar to each other, except their magnitudes.

\subsection{Analytic solutions}
\begin{figure}
\begin{center}
\includegraphics[width=10.5cm]{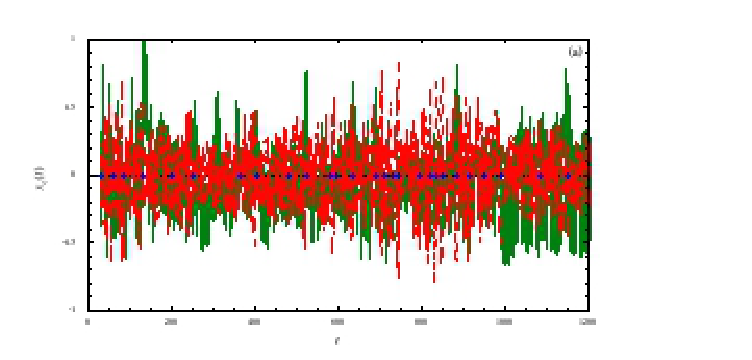}
\includegraphics[width=10.5cm]{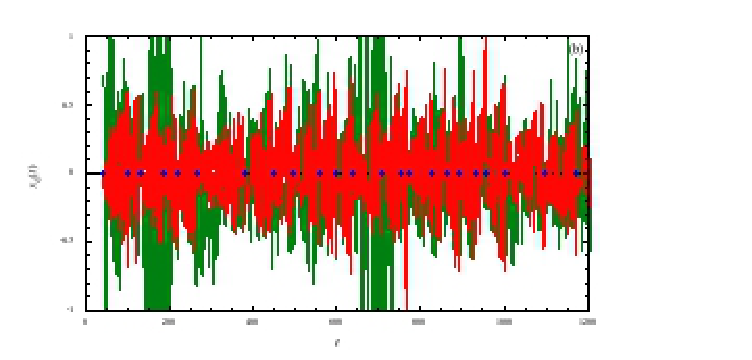}
\end{center}
\caption{(Color online) A plot of the fluctuations $x_{\alpha}(t)$ versus time $t$, where (a) $x_c$ and (b) $x_d$. The solid lines indicate the numerical results obtained by using the data observed in Japan and the dotted lines the numerical results obtained analytically by using Eq. (\ref{drf}). The fitting values of  $\gamma_{\alpha}$ is listed in Table \ref{table-1}. The details are the same as in Fig. \ref{jp1}.}
\label{jp12}
\end{figure}
\begin{figure}
\begin{center}
\includegraphics[width=10.5cm]{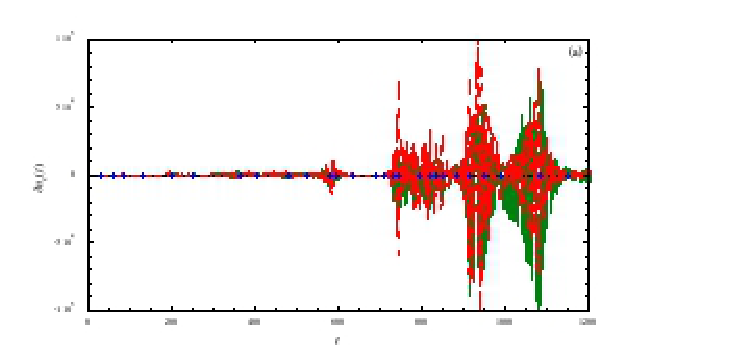}
\includegraphics[width=10.5cm]{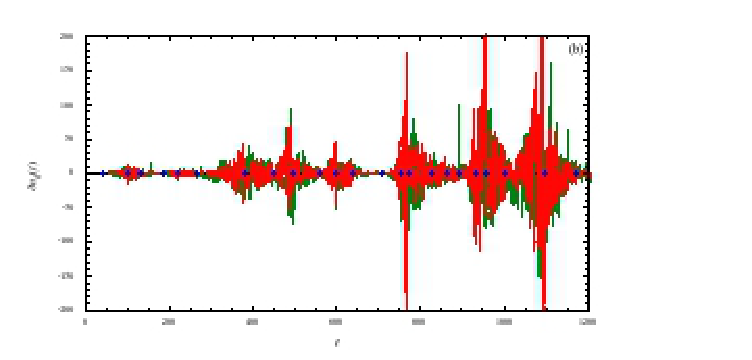}
\end{center}
\caption{(Color online) A plot of the fluctuations $\delta n_{\alpha}(t)$ observed in Japan versus time $t$, where (a) $\delta n_c(t)$ and (b) $\delta n_d(t)$. The details are the same as in Fig. \ref{jp12}.}
\label{jp13}
\end{figure}
In this subsection, we find the analytic solutions for $x_{\alpha}(t)$ and $\eta_{\alpha}(t)$ in terms of $\xi_{\alpha}(t)$. From Eq. (\ref{nome}), one can first write $\eta_{\alpha}(t)$ in terms of $x_{\alpha}(t)$ as
\begin{equation}
\eta_{\alpha}(t)=(\xi_{\alpha}(t)-\gamma_{\alpha})(x_{\alpha}(t)+1)=\frac{dx_{\alpha}(t)}{dt},\label{eta}
\end{equation}
where $\langle \eta_{\alpha}(t)\rangle=0$. Here we note that $\eta_{\alpha}(t)$ is not random in time since it also contains $x_{\alpha}(t)$. One can then formally solve Eq. (\ref{eta}) for $x_{\alpha}(t)$ to give
\begin{equation}
x_{\alpha}(t)=(1+x_{\alpha}(t_{\alpha}))e^{\int_{t_{\alpha}}^t(\xi_{\alpha}(s)-\gamma_{\alpha})ds}-1, \label{drf}
\end{equation}
which is combined with Eq. (\ref{eta}) to obtain
\begin{equation}
\eta_{\alpha}(t)=(1+x_{\alpha}(t_{\alpha}))(\xi_{\alpha}(t)-\gamma_{\alpha})e^{\int_{t_{\alpha}}^t(\xi_{\alpha}(s)-\gamma_{\alpha})ds}. \label{etasol}
\end{equation}
Thus, it turns out from Eqs. (\ref{drf}) and (\ref{etasol}) that $x_{\alpha}(t)$ and $\eta_{\alpha}(t)$ are uniquely determined by $\xi_{\alpha}(t)$. Here we note that $\eta_{\alpha}(t)$ has a term linearly proportional to $\xi_{\alpha}(t)$. As discussed before, this is a main reason why the dynamics of $\eta_{\alpha}(t)$ seems to be rather random compared to that of $x_{\alpha}(t)$, although $\eta_{\alpha}(t)$ is also governed by the exponential decay $e^{-\gamma_{\alpha}t}$ through $x_{\alpha}(t)$. Since the origin of the randomness in time for $\eta_{\alpha}(t)$ and $x_{\alpha}(t)$ results from that of $\xi_{\alpha}(t)$, it is reasonable to assume that $\xi_{\alpha}(t)$ is a Gaussian white noise with zero mean and satisfies 
\begin{eqnarray}
\langle \xi_{\alpha}(t)N_{\alpha}(t_{\alpha})\rangle&=&\langle \xi_{\alpha}(t)\rangle=0, \label{xi0}\\
\langle \xi_{\alpha}(t)\xi_{\alpha}(t')\rangle&=&2\gamma_{\alpha}\delta(t-t'). \label{xi}
\end{eqnarray}
Once the value of $\gamma_{\alpha}$ is known, therefore, $x_{\alpha}(t)$ and $\eta_{\alpha}(t)$ can be obtained numerically through Eqs. (\ref{drf}) and (\ref{etasol}), respectively. However, we note here that their dynamics are also governed by the exponential decay given by $e^{-\gamma t}$ even though $\xi_{\alpha}(t)$ is assumed to be random in time. Such a non-equilibrium behavior is easily seen in Fig. \ref{jp9}. The numerical behaviors similar to those discussed above are also shown to be seen for the data observed in various countries all over the world.

\begin{figure}
\begin{center}
\includegraphics[width=13.2cm]{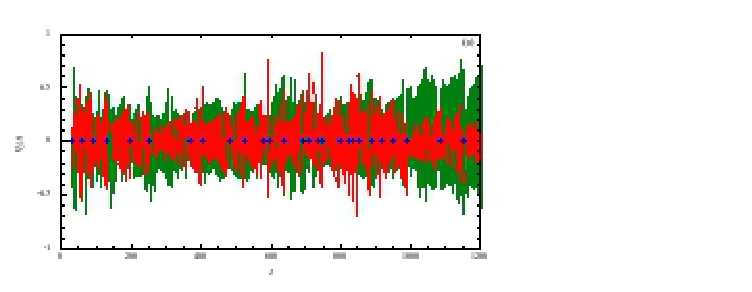}
\includegraphics[width=12.4cm]{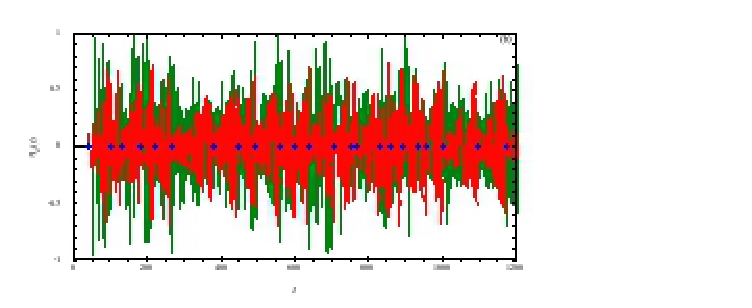}
\end{center}
\caption{(Color online) A plot of the stochastic forces $\eta_{\alpha}(t)$ versus time $t$, where (a) $\eta_c$ and (b) $\eta_d$. The solid lines indicate the numerical results obtained by using the data observed in Japan and the dotted lines the numerical results obtained analytically by using Eq. (\ref{etasol}).  The details are the same as in Fig. \ref{jp12}.}
\label{jp14}
\end{figure}
\begin{figure}
\begin{center}
\includegraphics[width=13.2cm]{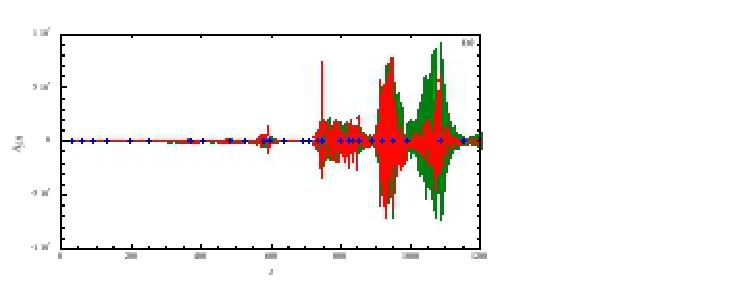}
\includegraphics[width=12.4cm]{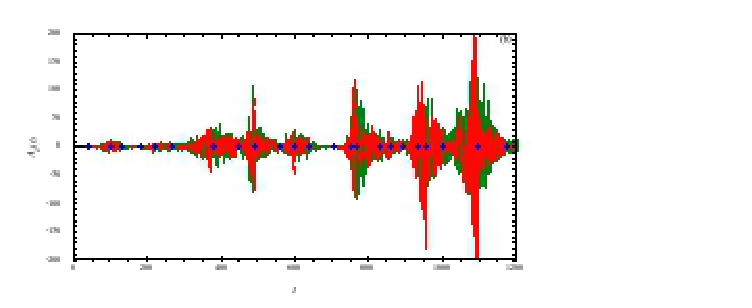}
\end{center}
\caption{(Color online) A plot of the stochastic forces $\Lambda_{\alpha}(t)$ observed in Japan versus time $t$, where (a) $\Lambda_c(t)$ and (b) $\Lambda_d(t)$. The dotted lines indicate the numerical results obtained by using Eq. (\ref{etasol}). The details are the same as in Fig. \ref{jp14}.}
\label{jp15}
\end{figure}
We next calculate the average quantities analytically and show that Eq. (\ref{avome}) is recovered based on the assumption for $\xi_{\alpha}(t)$. By using Eq. (\ref{Ome}), one can easily solve Eq. (\ref{Meq}) to give
\begin{equation}
N_{\alpha}(t)=N_{\alpha}(t_{\alpha})\exp\left[\lambda_{\alpha 0} (t-t_{\alpha})+\int_{t_{\alpha}}^t \xi_{\alpha}(s)ds\right]. \label{Nt}
\end{equation}
In order to calculate the average $\langle N_{\alpha}(t)\rangle$, it is convenient to employ the cumulant expansion method proposed by Kubo \cite{kubo63}. In fact, expanding Eq. (\ref{Nt}) in powers of $\xi_{\alpha}$ and taking the average of it, we find
\begin{eqnarray}
n_{\alpha}(t)&=&n_{\alpha}(t_{\alpha}) e^{\lambda_{\alpha 0}(t-t_{\alpha})}[1\nonumber\\
&+&\frac{1}{2}\int_{t_{\alpha}}^t ds\int_{t_{\alpha}}^t ds' \langle\xi_{\alpha}(s)\xi_{\alpha}(s')\rangle+\cdots], \label{Nex}
\end{eqnarray}
where we have used the fact that $\xi_{\alpha}(t)$ is statistically independent of $N_{\alpha}(t_{\alpha})$. Taking the logarithms of both sides of Eq. (\ref{Nex}) and expanding it in powers of $\xi_{\alpha}$, we finally obtain
\begin{eqnarray}
n_{\alpha}(t)
&=&n_{\alpha}(t_{\alpha})e^{[\lambda_{\alpha 0} (t-t_{\alpha})
+\frac{1}{2}\int_{t_{\alpha}}^t ds\int_{t_{\alpha}}^t ds' \langle\xi_{\alpha}(s)\xi_{\alpha}(s')\rangle]}\nonumber\\
&=&n_{\alpha}(t_{\alpha})e^{\lambda_{\alpha}(t-t_{\alpha})}, \label{Nts}
\end{eqnarray}
where the renormalized coefficient $\lambda_{\alpha}$ is given by
\begin {equation}
\lambda_{\alpha}=\lambda_{\alpha 0}+\gamma_{\alpha}. \label{rgdc}
\end {equation}
Here the higher order terms in $\xi_{\alpha}$ can be written in terms of cumulants of $\xi_{\alpha}(t)$ and exactly disappear because $\xi_{\alpha}(t)$ is a Gaussian white noise. Thus, Eq. (\ref{dnum}) is safely recovered, where the bare rate $\lambda_{\alpha 0}$ is turned out to be enhanced by the multiplicative stochastic noise since $\gamma_{\alpha}$ is a positive constant.  Hence the growth process is accelerated, while the decay process slows down. Similarly to the derivation of Eq. (\ref{Nts}), one can easily find
\begin{eqnarray}
\langle N_{\alpha}(t)N_{\alpha}(t')\rangle&=&n_{\alpha}(t)n_{\alpha}(t')[1+\langle x_{\alpha}(t)x_{\alpha}(t')\rangle],\label{NtNt'}\\
\langle x_{\alpha}(t)x_{\alpha}(t')\rangle&=&\chi_{\alpha}e^{2\gamma_{\alpha}[t-t_{\alpha}+\theta(t-t')(t'-t)]}-1,\label{x2}\\
\langle \Lambda_{\alpha}(t)\Lambda_{\alpha}(t')\rangle&=&n_{\alpha}(t)n_{\alpha}(t')\langle \eta_{\alpha}(t)\eta_{\alpha}(t')\rangle,\label{Rtt'}\\
\langle \eta_{\alpha}(t)\eta_{\alpha}(t')\rangle&=&2\gamma_{\alpha}\chi_{\alpha}e^{2\gamma_{\alpha}(t-t_{\alpha})}\delta(t-t'),
\label{etatt'}\\
\langle \eta_{\alpha}(t)\rangle&=&\langle \eta_{\alpha}(t)x_{\alpha}(t_{\alpha})\rangle=0, \label{x0}
\end{eqnarray}
where $\chi_{\alpha}=1+\langle x_{\alpha}(t_{\alpha})^2\rangle$. Here $\theta(t)$ is a step function which satisfies $\theta(t)=1$ for $t\geq0$ and $\theta(t)=0$ for $t<0$. By using Eqs. (\ref{Ome}) and (\ref{Nt}), one can also obtain
\begin{eqnarray}
\langle \Omega_{\alpha}(t)\rangle&=&\gamma_{\alpha}n_{\alpha}(t),  \label{Ots}\\
\langle \Omega_{\alpha}(t)\Omega_{\alpha}(t')\rangle&=&2\gamma_{\alpha}\langle N_{\alpha}(t)^2\rangle\delta(t-t')\nonumber\\&+&3\gamma_{\alpha}^2\langle N_{\alpha}(t)N_{\alpha}(t')\rangle.\label{Ott'}
\end{eqnarray}
\begin{figure}
\begin{center}
\includegraphics[width=11.1cm]{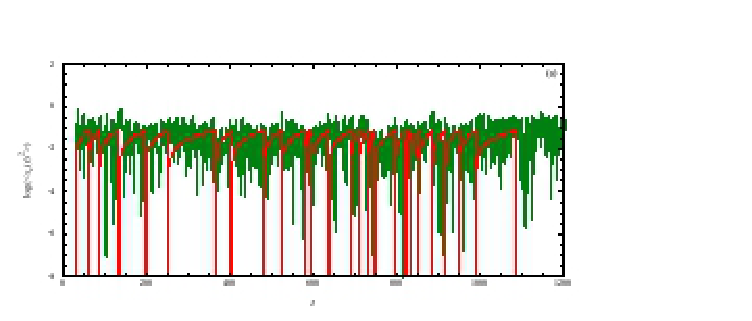}
\includegraphics[width=10.3cm]{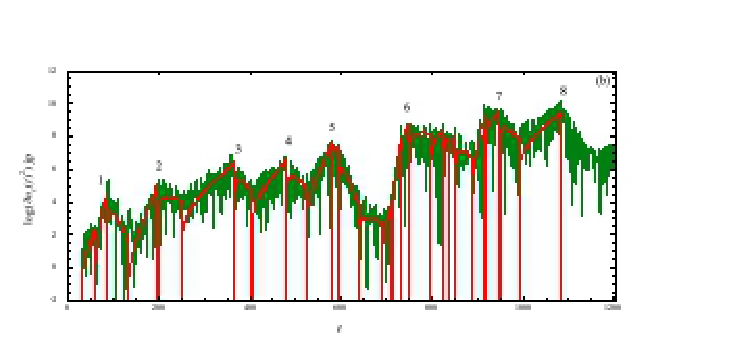}
\end{center}
\caption{(Color online) A log plot of the variance versus time $t$. (a) The solid lines indicates the analytic results for $<x_c(t)^2>$ given by Eq. (\ref{vx2}) and the fluctuating lines the observed data for $x_c(t)^2$ and (b) the solid lines indicates the analytic results for $<\delta n_c(t)^2>(=n_c(t)^2<x_c(t)^2>)$ and the fluctuating lines the observed data for $\delta n_c(t)^2$. The details are the same as in Fig. \ref{jp11}.}
\label{jp16}
\end{figure}
\begin{figure}
\begin{center}
\includegraphics[width=11.0cm]{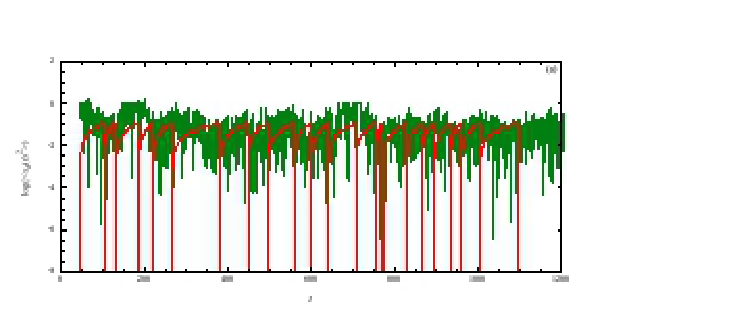}
\includegraphics[width=11.0cm]{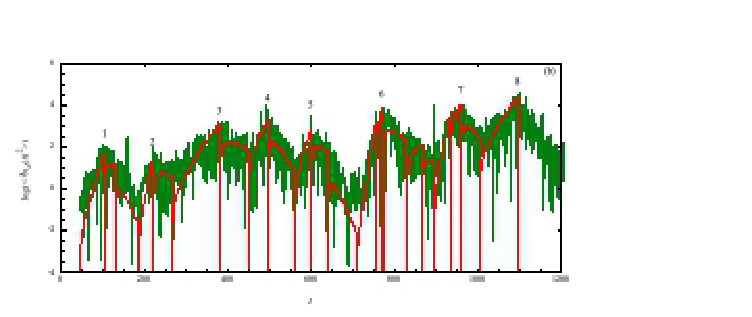}
\end{center}
\caption{(Color online) A log plot of the variance versus time $t$. (a) The solid lines indicates the analytic results for $<x_d(t)^2>$ given by Eq. (\ref{vx2}) and the fluctuating lines the observed data for $x_d(t)^2$ and (b) the solid lines indicates the analytic results for $<\delta n_d(t)^2>(=n_d(t)^2<x_d(t)^2>)$ and the fluctuating lines the observed data for $\delta n_d(t)^2$. The details are the same as in Fig. \ref{jp11}.}
\label{jp17}
\end{figure}
Thus, Eq. (\ref{avome}) is consistently recovered under the assumption that $\xi_{\alpha}(t)$ is a Gaussian white noise. Use of Eq. (\ref{x0}) also leads to
\begin{equation}
\langle \Lambda_{\alpha}(t)\rangle=\langle \Lambda_{\alpha}(t)N_{\alpha}(t_{\alpha})\rangle=0. \label{L0}
\end{equation}

Finally, we also discuss the average of $\delta f_{\alpha}(t)$. By using Eq. (\ref{yt}), one can easily find 
\begin{eqnarray}
&&\langle \delta f_{\alpha}(t)\delta f_{\alpha}(t')\rangle
=\langle \delta f_{\alpha}(t_{\alpha})^2\rangle\nonumber\\
&+&\int_{t_{\alpha}}^t ds\int_{t_{\alpha}}^{t'} ds' n_{\alpha}(s)n_{\alpha}(s')\langle x_{\alpha}(s)x_{\alpha}(s')\rangle. \label{yave}
\end{eqnarray}
Use of Eqs. (\ref{x2}) and (\ref{yave}) then leads to
\begin{equation}
\langle \delta f_{\alpha}(t)^2\rangle=f_{\alpha}(t)^2\langle y_{\alpha}(t)^2\rangle \label{fave2}
\end{equation}
with
\begin{eqnarray}
\langle y_{\alpha}(t)^2\rangle&=&\bigg(\frac{f_{\alpha}(t_{\alpha})}{f_{\alpha}(t)}\bigg)^2\langle y_{\alpha}(t_{\alpha})^2\rangle\nonumber\\
&+&\bigg(\frac{n_{\alpha}(t_{\alpha})}{f_{\alpha}(t)}\bigg)^2\bigg[\frac{\chi_{\alpha}(e^{2(\lambda_{\alpha}+\gamma_{\alpha})(t-t_{\alpha})}-1)}{(\lambda_{\alpha}+\gamma_{\alpha})(\lambda_{\alpha}+2\gamma_{\alpha})}\nonumber\\
&-&\frac{2\chi_{\alpha}(e^{\lambda_{\alpha}(t-t_{\alpha})}-1)}{\lambda_{\alpha}(\lambda_{\alpha}+2\gamma_{\alpha})}-\frac{(e^{\lambda_{\alpha}(t-t_{\alpha})}-1)^2}{\lambda_{\alpha}^2}\bigg]. \nonumber\\
\label{yave2}
\end{eqnarray}
In Sec. IV, the variances $\langle \delta n_{\alpha}(t)^2\rangle$ and $\langle \delta f_{\alpha}(t)^2\rangle$ are numerically calculated by using the fitting values of unknown parameters. Then, they are directly compared with the fluctuations $\delta n_{\alpha}(t)^2$ and $\delta f_{\alpha}(t)^2$, which are obtained from the observed data. This is because the average quantities of the observed data are not available since the system is not in equilibrium.

\subsection{A Langevin-type equation}
In this subsection, we first derive the Langevin-type equation for $N_{\alpha}(t)$, starting from the stochastic equation (\ref{Meq}). 
In order to find such an equation, it is convenient to employ the time-convolutionless projection-operator method proposed by Tokuyama-Mori \cite{toku75,toku76}. In fact, the derivation of the Langevin-type equation and the corresponding master equation for the probability distribution function from the stochastic equation with the multiplicative stochastic force has been already discussed in the previous papers \cite{toku80,toku81}. Let $P(n_{\alpha},t)$ denote the probability distribution function for $N_{\alpha}(t)$ to have a value $n_{\alpha}$ at time $t$. Following the same approach as that shown in Refs. \cite{toku80,toku81}, one can transform Eq. (\ref{Meq}) into the Fokker-Plank type equation for $P(n_{\alpha},t)$
\begin{equation}
\frac{\partial}{\partial t}P(n_{\alpha},t)=\frac{\partial}{\partial n_{\alpha}}\left[-\lambda_{\alpha}n_{\alpha}+\gamma_{\alpha}\frac{\partial}{\partial n_{\alpha}}n_{\alpha}^2\right]P(n_{\alpha},t), \label{FPeq}
\end{equation}
and the corresponding Langevin-type equation for $N_{\alpha}(t)$
\begin{equation}
\frac{d}{dt}N_{\alpha}(t)=\lambda_{\alpha}N_{\alpha}(t)+R_{\alpha}(t),\label{Leq}
\end{equation}
where $R_{\alpha}(t)$ is an additive-type stochastic force and satisfies
\begin{equation}
\langle R_{\alpha}(t)\rangle=\langle R_{\alpha}(t)N_{\alpha}(t_{\alpha})\rangle=0, \label{fd0}
\end{equation}
\begin{equation}
\langle R_{\alpha}(t)R_{\alpha}(t')\rangle=2\gamma_{\alpha}\langle N_{\alpha}(t)^2\rangle\delta(t-t').\label{Rtt'}
\end{equation}
\begin{figure}
\begin{center}
\includegraphics[width=11.0cm]{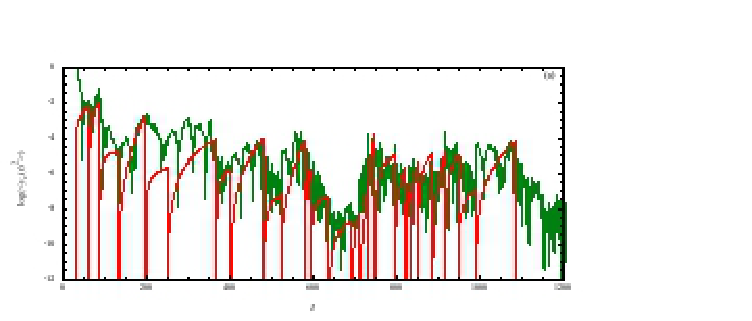}
\includegraphics[width=11.0cm]{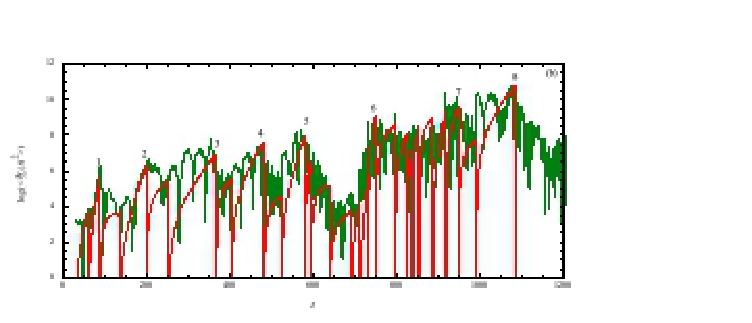}
\end{center}
\caption{(Color online) A log plot of the variance versus time $t$. (a) The solid lines indicates the analytic results for $<y_c(t)^2>$ and the fluctuating lines the observed data for $y_c(t)^2$ and (b) the solid lines indicates the analytic results for $<\delta f_c(t)^2>$ and the fluctuating lines the observed data for $\delta f_c(t)^2$}
\label{jp18}
\end{figure}
\begin{figure}[t]
\begin{center}
\includegraphics[width=11.0cm]{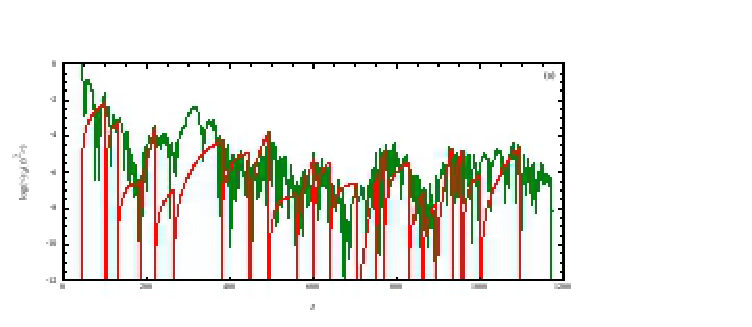}
\includegraphics[width=11.0cm]{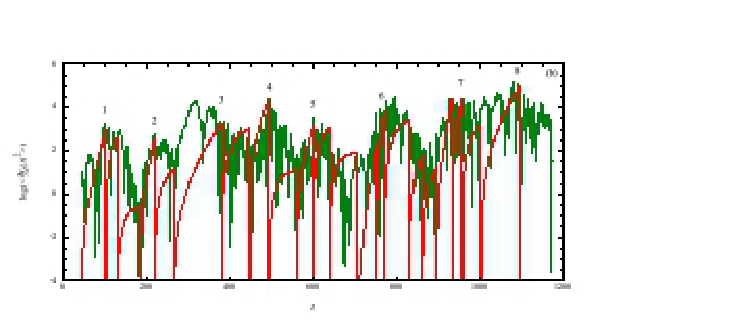}
\end{center}
\caption{(Color online) A log plot of the variance versus time $t$. (a) The solid lines indicate the analytic results for $<y_d(t)^2>$ and the fluctuating lines the observed data for $y_d(t)^2$ and (b) the solid lines indicate the analytic results for $<\delta f_d(t)^2>$ and the fluctuating lines the observed data for $\delta f_d(t)^2$}
\label{jp19}
\end{figure}
Thus, it turns out that the stochastic force $R_{\alpha}(t)$ is identical to the function $\Lambda_{\alpha}(t)$ introduced by Eq. (\ref{nome}). In order to derive Eq. (\ref{FPeq}), we have employed the same manner as that discussed in Ref. \cite{toku81}. In fact, starting from Eq. (\ref{Meq}), one can first derive the master equation for $P(n_{\alpha},t)$ exactly, which is given by $(\partial/\partial t)P(n_{\alpha},t)=M(n_{\alpha},t)P(n_{\alpha},t)$ with the master operator $M(n_{\alpha},t)$. Then, $M(n_{\alpha},t)$ is shown to be written in terms of cumulants of $\Omega_{\alpha}(n_{\alpha},t)(=n_{\alpha}\xi_{\alpha}(t))$. Since $\xi_{\alpha}(t)$ is assumed to be a Gaussian white noise and $\langle\Omega_{\alpha}(n_{\alpha},t)\rangle=0$, all higher than second-order cumulants of $\Omega_{\alpha}(n_{\alpha},t)$ are then shown to exactly vanish. Thus, one can obtain Eq. (\ref{FPeq}). It turns out from Eqs. (\ref{Meq}) and (\ref{Leq}) that the multiplicative stochastic force $\Omega_{\alpha}(t)$ is separated into the systematic part $\gamma_{\alpha}N_{\alpha}(t)$ and the additive-type stochastic force $R_{\alpha}(t)$ which coincides with the function $\Lambda_{\alpha}(t)$ introduced in Eq. (\ref{nome}). Here we note that although the correlation function of the stochastic force $R_{\alpha}(t)$ is $\delta$-correlated in time, it still depends on time through the variance $\langle N_{\alpha}(t)^2\rangle$. We also mention that since the second derivative term in Eq. (\ref{FPeq}) contains a nonlinear term $n_{\alpha}^2$, Eq. (\ref{FPeq}) is not an usual Fokker-Planck equation, where the second derivative term contains a constant. Equation (\ref{Leq}) is now solved for $\delta n_{\alpha}(t)$ to give
\begin{equation}
\delta n_{\alpha}(t)=e^{\lambda(t-t_{\alpha})}\delta n_{\alpha}(t_{\alpha})+\int_{t_{\alpha}}^t e^{\lambda(t-s)}R_{\alpha}(s)ds. \label{dns}
\end{equation}
Hence it turns out that the fluctuations $\delta n_{\alpha}(t)$ can be written in terms of $R_{\alpha}(t)$. 

The Fokker-Planck equation (\ref{FPeq}) is useful to calculate the average $\langle N_{\alpha}(t)^{\nu}\rangle(=\int n_{\alpha}^{\nu}P(n_{\alpha},t)dn_{\alpha})$ analytically, where $\nu(\geq 1)$ is an integer. Use of Eq. (\ref{FPeq}) then leads to
\begin{equation}
\frac{d}{dt}\langle N_{\alpha}(t)^{\nu}\rangle=\nu\left[\lambda_{\alpha}+(\nu-1)\gamma_{\alpha}\right]\langle N_{\alpha}(t)^{\nu}\rangle.\label{dnell}
\end{equation}
This is easily solved to give
\begin{equation}
\langle N_{\alpha}(t)^{\nu}\rangle=n_{\alpha}(t)^{\nu}e^{\nu(\nu-1)\gamma_{\alpha}(t-t_{\alpha})}\langle [1+x_{\alpha}(t_{\alpha})]^{\nu}\rangle, \label{nell}
\end{equation}
which also leads to
\begin{equation}
\langle [1+x_{\alpha}(t)]^{\nu}\rangle=e^{\nu(\nu-1)\gamma_{\alpha}(t-t_{\alpha})}\langle [1+x_{\alpha}(t_{\alpha})]^{\nu}\rangle. \label{nellx}
\end{equation}
When $\nu=2$, Eq. (\ref{nellx}) reduces to Eq. (\ref{x2}) at $t'=t$.

\begin{figure}
\begin{center}
\includegraphics[width=11.0cm]{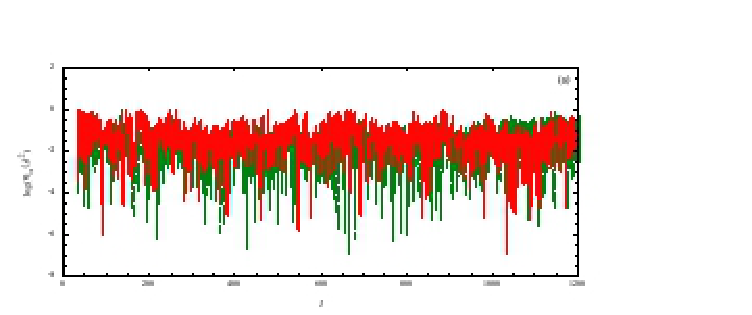}
\includegraphics[width=11.0cm]{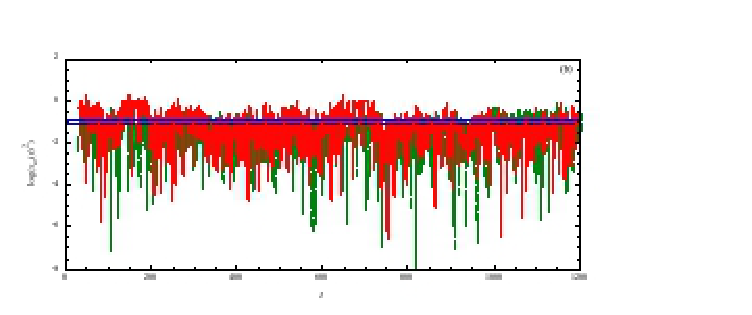}
\includegraphics[width=11.0cm]{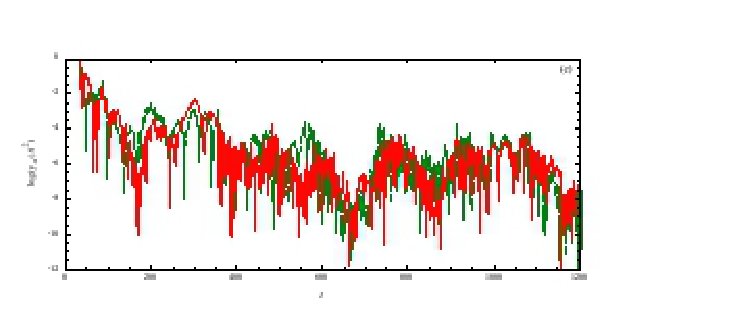}
\end{center}
\caption{(Color online) A log plot of the fluctuations versus time $t$. (a) $\eta_{\alpha}(t)^2$ , (b) $x_{\alpha}(t)^2$, and (c) $y_{\alpha}(t)^2$. The fluctuating solid lines indicate the results for death and the fluctuating dashed lines for case. The horizontal solid line indicates the mean square $Q_d^2$ and the horizontal dashed line $Q_c^2$. The details are the same as in Fig. \ref{jp7}.}
\label{jp20}
\end{figure}
\begin{figure}
\begin{center}
\includegraphics[width=11.0cm]{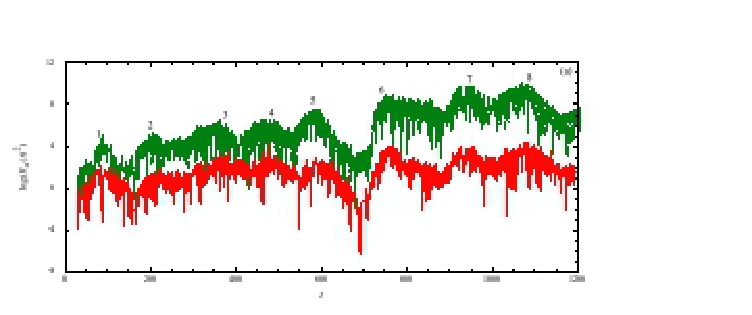}
\includegraphics[width=11.0cm]{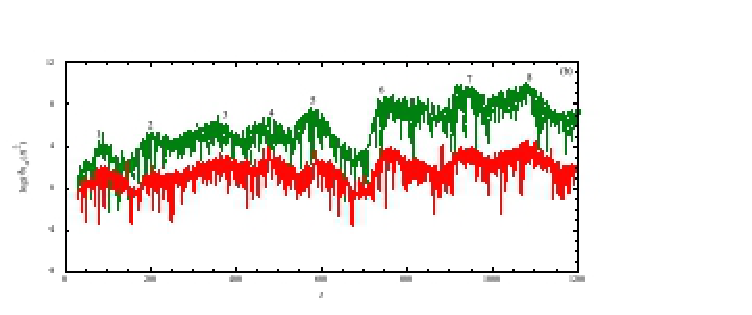}
\includegraphics[width=11.0cm]{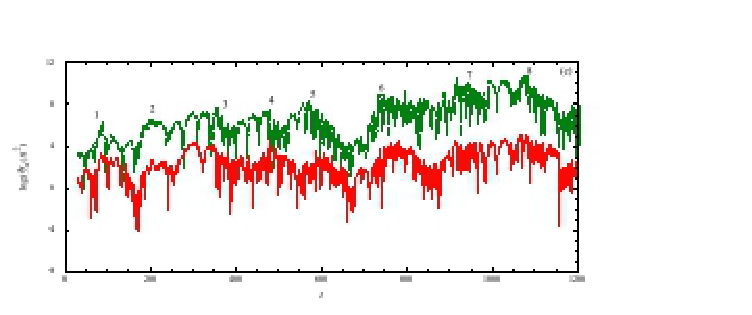}
\end{center}
\caption{(Color online) A log plot of the fluctuations versus time $t$. (a) $R_{\alpha}(t)^2$ , (b) $\delta n_{\alpha}(t)^2$, and (c) $\delta f_{\alpha}(t)^2$. The details are the same as in Fig. \ref{jp20}.}
\label{jp21}
\end{figure}
\begin{figure}
\begin{center}
\includegraphics[width=11.0cm]{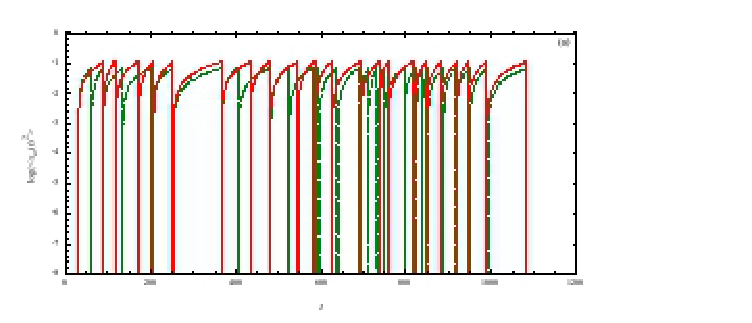}
\includegraphics[width=11.0cm]{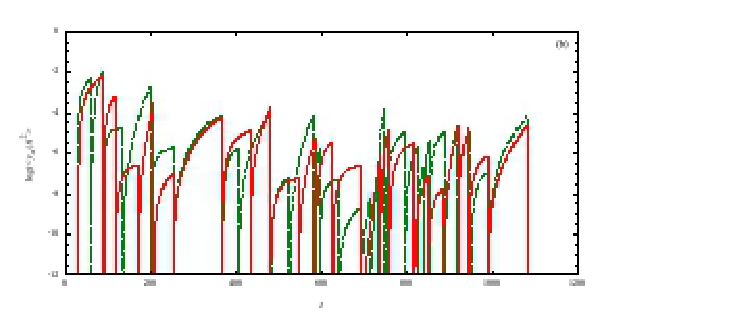}
\includegraphics[width=11.0cm]{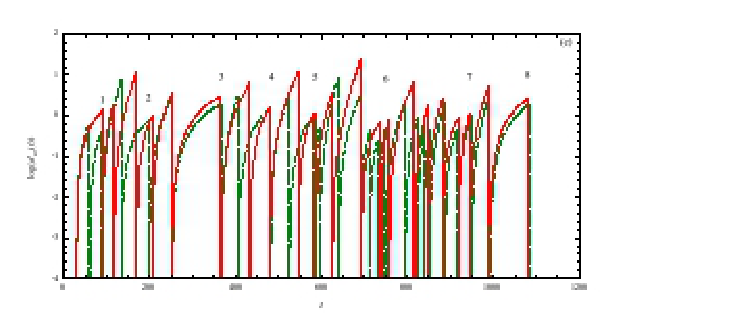}
\end{center}
\caption{(Color online) A log plot of the analytic results versus time $t$, where (a) $<x_{\alpha}(t)^2>$, (b) $<y_{\alpha}(t)^2>$, and (c) $d_{\alpha}(t)$. The solid lines indicate the results for death and the dashed lines for case. The details are the same as in Fig. \ref{jp7}.}
\label{jp22}
\end{figure}
\begin{figure}
\begin{center}
\includegraphics[width=11.0cm]{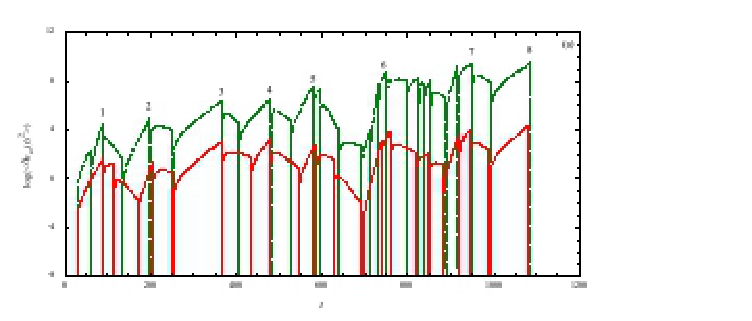}
\includegraphics[width=11.0cm]{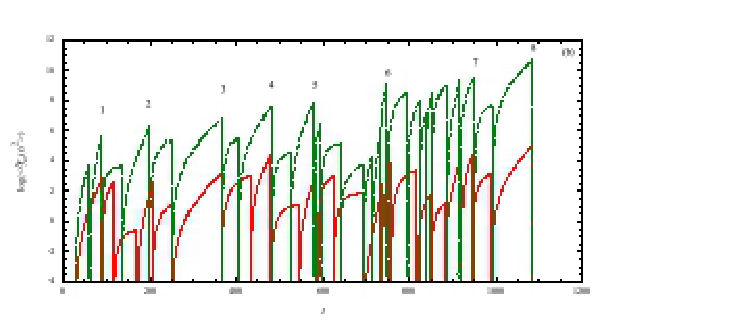}
\includegraphics[width=11.0cm]{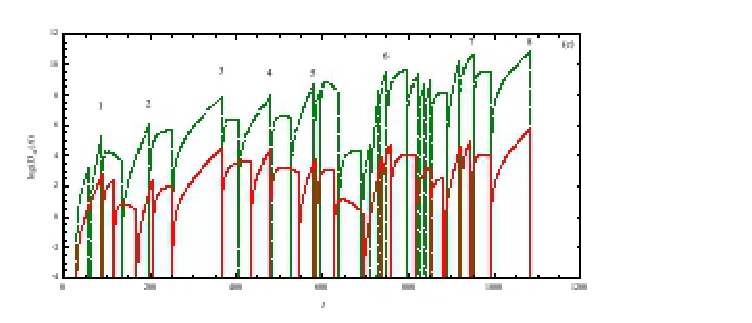}
\end{center}
\caption{(Color online) A log plot of the analytic results versus time $t$, where (a) $<\delta n_{\alpha}(t)^2>$, (b) $<\delta f_{\alpha}(t)^2>$, and (c) $D_{\alpha}(t)$. The details are the same as in Fig. \ref{jp22}.}
\label{jp23}
\end{figure}
\begin{figure}
\begin{center}
\includegraphics[width=9.5cm]{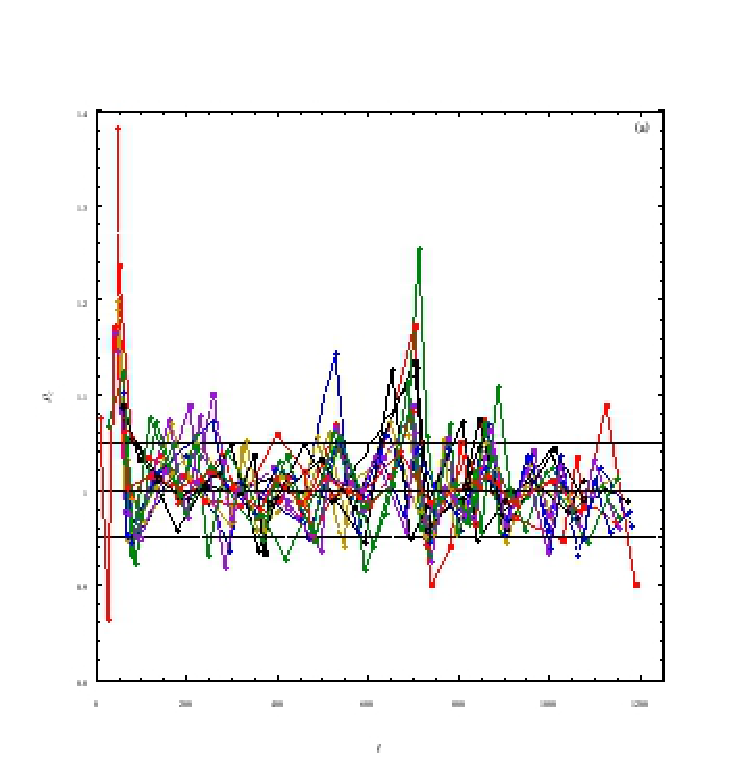}
\includegraphics[width=9.5cm]{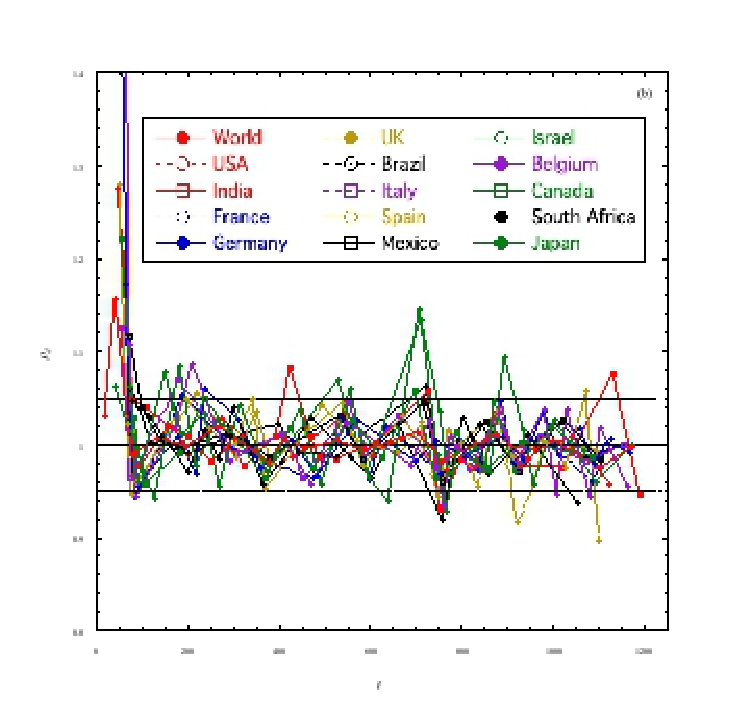}
\end{center}
\caption{(Color online) A log plot of the infection rate $\rho_{\alpha}$ versus time $t$, where (a) $\rho_c$ and (b) $\rho_d$. The symbols indicate the fitting values of $\rho_{\alpha}$ in different countries and the various lines a guide to eyes. The horizontal solid lines indicate $\rho_{\alpha}=1.00$, the horizontal dotted lines $\rho_{\alpha}=1.05$, and the horizontal dashed lines $\rho_{\alpha}=0.95$. The starting time $t=1$ is adjusted to be January 15, 2020 when the first case data were observed.}
\label{jp24}
\end{figure}
\begin{figure}
\begin{center}
\includegraphics[width=9.5cm]{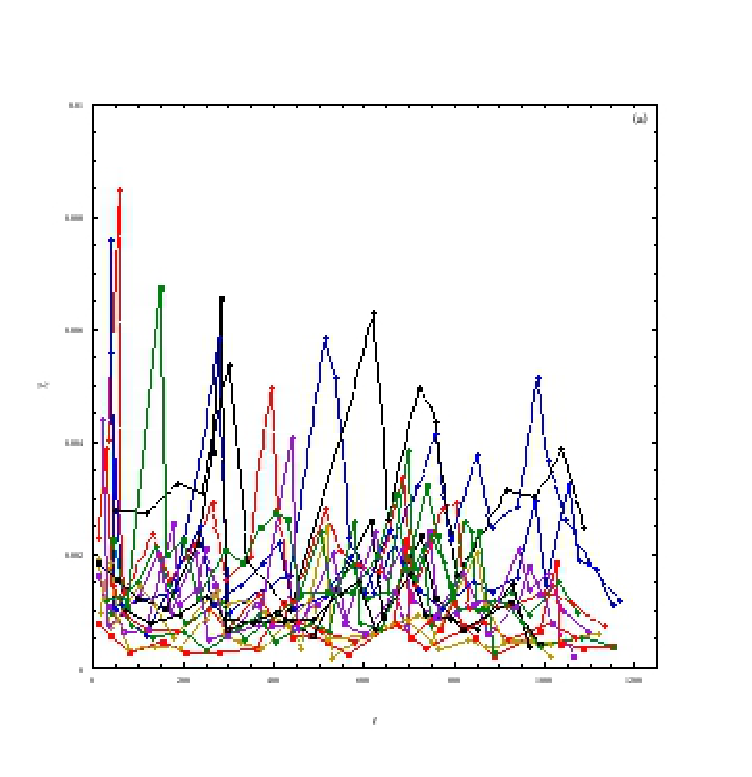}
\includegraphics[width=9.5cm]{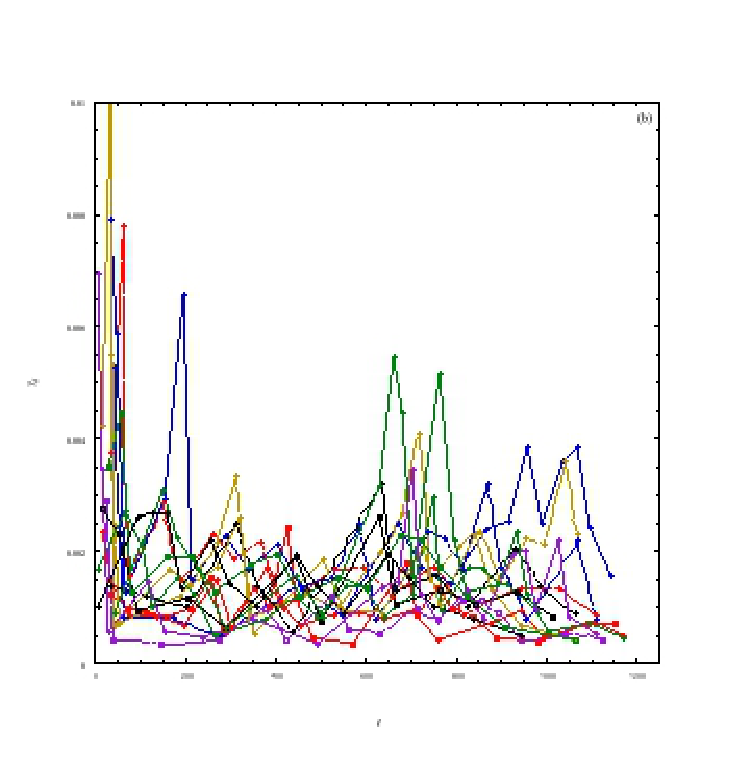}
\end{center}
\caption{(Color online) A log plot of  $\gamma_{\alpha}$ versus time $t$, where (a) $\gamma_c$ and (b) $\gamma_d$. The symbols indicate the fitting values of $\gamma_{\alpha}$ in different countries. The details are the same as in Fig. \ref{jp24}.}
\label{jp25}
\end{figure}
Finally, we also discuss the stochastic equation for $F_{\alpha}(t)$ and the corresponding Fokker-Planck type equation. Let $P(f_{\alpha},t)$ denote the probability distribution function for $F_{\alpha}(t)$ to have a value $f_{\alpha}$ at time $t$. Since $F_{\alpha}(t)$ can be written as
\begin{equation}
\frac{d}{dt}F_{\alpha}(t)=n_{\alpha}(t)+\delta n_{\alpha}(t),  \label{dFndt}
\end{equation}
similarly to the derivation of Eq. (\ref{FPeq}), one can also obtain the Fokker-Planck type equation for $P(f_{\alpha},t)$ as
\begin{equation}
\frac{\partial}{\partial t}P(f_{\alpha},t)=\frac{\partial}{\partial f_{\alpha}}\left[-n_{\alpha}(t)+D_{\alpha}(t)\frac{\partial}{\partial f_{\alpha}}\right]P(f_{\alpha},t) \label{FPfeq}
\end{equation}
with the diffusion-like coefficient 
\begin{equation}
D_{\alpha}(t)=\int_{t_{\alpha}}^t \langle \delta n_{\alpha}(t)\delta n_{\alpha}(s)\rangle ds.\label{cD}
\end{equation}
Here we note that one can also write $D_{\alpha}(t)$ in terms of the correlation function of $R_{\alpha}(t)$ by using Eq. (\ref{dns}), although it becomes rather complicated because of triple integrals in time. Use of Eqs. (\ref{x2}) and (\ref{cD}) then leads to
\begin{eqnarray}
D_{\alpha}(t)&=&n_{\alpha}(t)^2d_{\alpha}(t),\\  \label{Dat}
d_{\alpha}(t)&=&\frac{\chi_{\alpha}}{\lambda_{\alpha}+2\gamma_{\alpha}}\big(e^{2\gamma_{\alpha}(t-t_{\alpha})}-e^{-\lambda_{\alpha}(t-t_{\alpha})}\big)\nonumber\\
&-&\frac{1}{\lambda_{\alpha}}\big(1-e^{-\lambda_{\alpha}(t-t_{\alpha})}\big). \label{da}
\end{eqnarray}
Similarly to Eq. (\ref{dnell}), from Eq. (\ref{FPfeq}), one can also find the equation for the average $\langle F_{\alpha}(t)^{\nu}\rangle(=\int f_{\alpha}^{\nu}P(f_{\alpha},t)df_{\alpha}$) as
\begin{eqnarray}
\frac{d}{dt}\langle F_{\alpha}(t)^{\nu}\rangle&=&\nu n_{\alpha}(t) \langle F_{\alpha}(t)^{\nu-1}\rangle\nonumber\\
&+&\nu(\nu-1)D_{\alpha}(t)\langle F_{\alpha}(t)^{\nu-2}\rangle.\label{ff}
\end{eqnarray}
By solving Eq. (\ref{ff}), one can then obtain Eq. (\ref{NF}) when $\nu=1$ and also Eq. (\ref{yave2}) when $\nu=2$.
By using Eq. (\ref{dns}), one can also write Eq. (\ref{yt}) in terms of $R_{\alpha}(t)$ as
\begin{eqnarray}
\delta f_{\alpha}(t)&=&\delta f_{\alpha}(t_{\alpha})+\frac{\delta n_{\alpha}(t_{\alpha})}{\lambda_{\alpha}}[e^{\lambda_{\alpha}(t-t_{\alpha})}-1]\nonumber\\
&+&\frac{1}{\lambda_{\alpha}}\int_{t_{\alpha}}^t [e^{\lambda_{\alpha}(t-s)}-1]R_{\alpha}(s)ds. \label{dfs}
\end{eqnarray}
Thus, not only the fluctuations $\delta n_{\alpha}(t)$ but also the fluctuations $\delta f_{\alpha}(t)$ can be written in terms of $R_{\alpha}(t)$ analytically. Hence it is understandable that the dynamics of the observed fluctuations for $\delta n_{\alpha}(t)$ and $\delta f_{\alpha}(t)$ are similar to those of the observed fluctuations for $R_{\alpha}(t)$ (see Figs. \ref{jp9} (a) and \ref{jp9} (c)). Here we note that use of Eqs. (\ref{Rtt'}) and (\ref{dfs}) easily leads to Eq. (\ref{yave2}).

Finally, we discuss the magnitudes of the fluctuations $\delta n_{\alpha}(t)$, $\delta f_{\alpha}(t)$, and $R_{\alpha}(t)$. From Eqs. (\ref{dns}) and (\ref{dfs}), one can estimate the magnitudes of $\delta n_{\alpha}(t)$ and $\delta f_{\alpha}(t)$ analytically as
\begin{eqnarray}
\delta n_{\alpha}(t)&\sim& \frac{R_{\alpha}(t)}{\lambda_{\alpha}},  \;\;x_{\alpha}(t)\sim \frac{\eta_{\alpha}(t)}{\lambda_{\alpha}}\label{On}\\
\delta f_{\alpha}(t)&\sim& \frac{R_{\alpha}(t)}{\lambda_{\alpha}^2}, \;\;y_{\alpha}(t)\sim \frac{\eta_{\alpha}(t)}{\lambda_{\alpha}^2}\bigg(\frac{n_{\alpha}(t)}{f_{\alpha}(t)}\bigg),\label{Of}
\end{eqnarray}
respectively. Since $|\lambda_{\alpha}|<1$ and $n_{\alpha}(t)/f_{\alpha}(t)\ll 1$ as shown in Fig. \ref{jp7} (b), we then find
\begin{eqnarray}
|\delta f_{\alpha}(t)|&>&|\delta n_{\alpha}(t)|>|R_{\alpha}(t)|, \label{flu1}\\
|x_{\alpha}(t)|&>&|\eta_{\alpha}(t)|\gg|y_{\alpha}(t)|,\label{flu2}
\end{eqnarray}
In fact, these relations are directly confirmed by comparing the observed data with each other (see Fig. \ref{jp9}). 

\section{Numerical solutions}
As discussed before, all the parameters contained in the present paper, except $\gamma_{\alpha}$, are shown to be found consistently well by fitting $f_{\alpha}(t)$ with the data observed in each communities. Since the value of $\gamma_{\alpha}$ is related to the stochastic force $\xi_{\alpha}(t)$ through Eq. (\ref{xi}), one can generate the random numbers for $\xi_{\alpha}(t)$ by choosing a appropriate value for $\gamma_{\alpha}$. The analytic forms of $x_{\alpha}(t)$ and $\eta_{\alpha}(t)$ are given by Eqs. (\ref{drf}) and (\ref{etasol}) as a function of $\xi_{\alpha}(t)$, respectively. Hence one can calculate their numerical values by using those random numbers. However, we should note here that the only one observed data are available in each country since the stochastic processes are irreversible in time. Hence it is rather difficult to fix the value of $\gamma_{\alpha}$ by comparing those numerical values with the observed data.

\subsection{A value of $\gamma_{\alpha}$}
In this subsection, we assume that the stochastic properties of $x_{\alpha}(t)$ should be the same for all stages in each community. 
In order to find the value of $\gamma_{\alpha}$ appropriately, therefore, it is convenient to use the variance $\langle x_{\alpha}(t)^2\rangle$, whose analytic form is given from Eq. (\ref{nellx}) (or Eq. (\ref{x2}) by
\begin{equation}
\langle  x_{\alpha}(t)^2\rangle=\chi_{\alpha}e^{2\gamma_{\alpha}(t-t_{\alpha})}-1.\label{vx2}
\end{equation}
At $i$th stage in the $\ell$th wave, $\langle  x_{\alpha}(t)^2\rangle$ starts to grow in time
from $\chi_{\alpha}-1$ at $t=t_{\alpha \ell}^{(i)}$ up to the maximum value at $t=t_{\alpha \ell}^{(i+1)}-1$, which is given by 
\begin{equation}
\langle  x_{\alpha}(t_{\alpha \ell}^{(i+1)}-1)^2\rangle=\chi_{\alpha}e^{2\gamma_{\alpha}(t_{\alpha \ell}^{(i+1)}-1-t_{\alpha})}-1.\label{Qcon}
\end{equation}
Then, it starts to decay in time at $t=t_{\alpha \ell}^{(i+1)}$ up to the minimum value at $t=t_{\alpha \ell}^{(i+2)}-1$. This jagged process is then repeated at each stage. In order to find the maximum value given by Eq. (\ref{Qcon}), one may use the root mean square $Q_{\alpha}$ of the observed data for $x_{\alpha}(t)$, which is shown by the bold horizontal line in Fig. \ref{jp9}. In fact, as is shown in Fig. \ref{jp10}, one can estimate the value of $Q_{\alpha}$ approximately for different communities, including prefectures and states. Here we note that although the different values must be caused by the complicated social conditions in each community, the inequality $Q_d>Q_c$ always holds for any communities since $|x_d(t)|> |x_c(t)|$ (or $n_c(t)> n_d(t)$). There exists an ambiguity for the fitting value of $Q_{\alpha}$ since the dynamics of $N_{\alpha}(t)$ is irreversible and the observed data is available only once. In order to find a reasonable value $Q_{\alpha}$, therefore, we take only the observed data for $x_{\alpha}(t)$ which satisfy the condition $|x_{\alpha}(t)|\leq 1$. Once the value of $Q_{\alpha}$ is found,  one may put $\langle  x_{\alpha}(t_{\alpha \ell}^{(i+1)}-1)^2\rangle=Q_{\alpha}^2$ and then use Eq. (\ref{Qcon}) to find the value of $\gamma_{\alpha}$ as
\begin{equation}
\gamma_{\alpha \ell}^{(i)}=\frac{\ln(1+Q_{\alpha}^2)-\ln(\chi_{\alpha})}{2(t_{\alpha \ell}^{(i+1)}-1-t_{\alpha \ell}^{(i)})}.\label{ga2}
\end{equation}
The values of $Q_{\alpha}$ and $\gamma_{\alpha}$ at each stage in Japan are listed in Table \ref{table-1}, where we have just set $x_{\alpha}(t_{\alpha})=0$ (or $\chi_{\alpha}=1$) for simplicity. In Fig. \ref{jp11}, the value of $\gamma_{\alpha}$ is also shown versus time $t_{\alpha}$. The time dependence of $\gamma_d$ is similar to that of $\gamma_c$. This situation is the same as that of $\rho_{\alpha}$ (see Fig. \ref{jp7} (a)). Hence we further confirm that there exists a strong correlation between the non-equilibrium stochastic process for case and that for death. 

\subsection{Numerical calculations of $x_{\alpha}(t)$ and $\eta_{\alpha}(t)$}
By generating the random numbers for $\xi_{\alpha}(t)$ under the fixed value of $\gamma_{\alpha}$ and using Eq. (\ref{drf}), one can now obtain the numerical results for $x_{\alpha}(t)$. In Figs. \ref{jp12} and \ref{jp13}, the numerical results for $x_{\alpha}(t)$ and $\delta n_{\alpha}(t)$ are thus shown versus time together with the observed data, respectively. Their time dependence also seems to be approximately well recovered. Similarly to $x_{\alpha}(t)$, one can also find the numerical results for $\eta_{\alpha}(t)$ by using Eq. (\ref{etasol}). In Figs. \ref{jp14} and \ref{jp15}, the numerical results for $\eta_{\alpha}(t)$ and $R_{\alpha}(t)$ are then plotted versus time together with the observed data, respectively. Their time dependence also seems to be approximately well recovered. 
\begin{figure}
\begin{center}
\includegraphics[width=10.5cm]{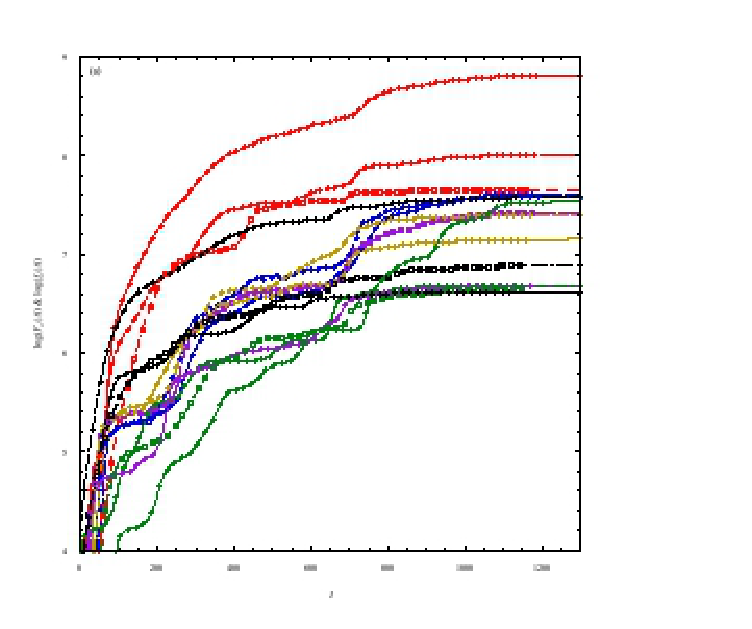}
\includegraphics[width=10.5cm]{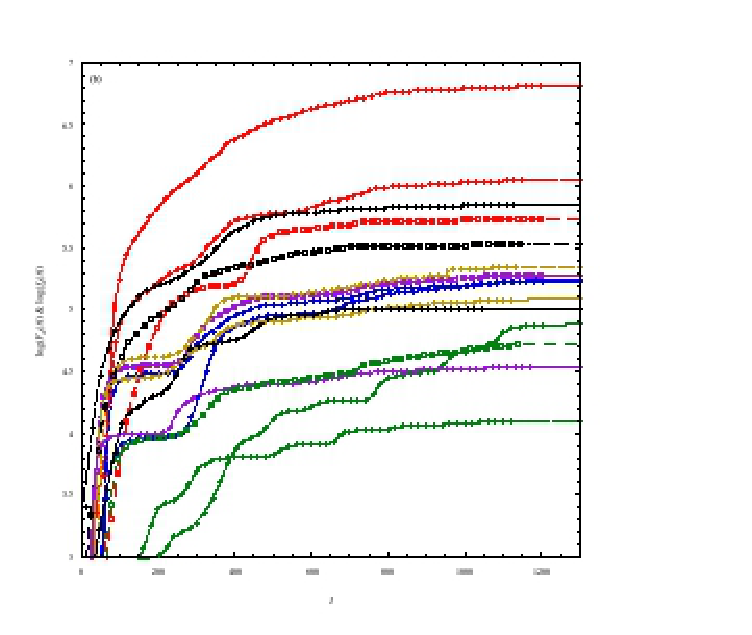}
\end{center}
\caption{(Color online) A log plot of $F_{\alpha}(t)$ and $f_{\alpha}(t)$ versus time $t$, where (a) case and (b) death. The symbols indicate the observed data for $F_{\alpha}(t)$ in different countries and the various solid lines the analytic results for $f_{\alpha}(t)$. The details are the same as in Fig. \ref{jp24}.}
\label{jp26}
\end{figure}
\begin{figure}
\begin{center}
\includegraphics[width=9.0cm]{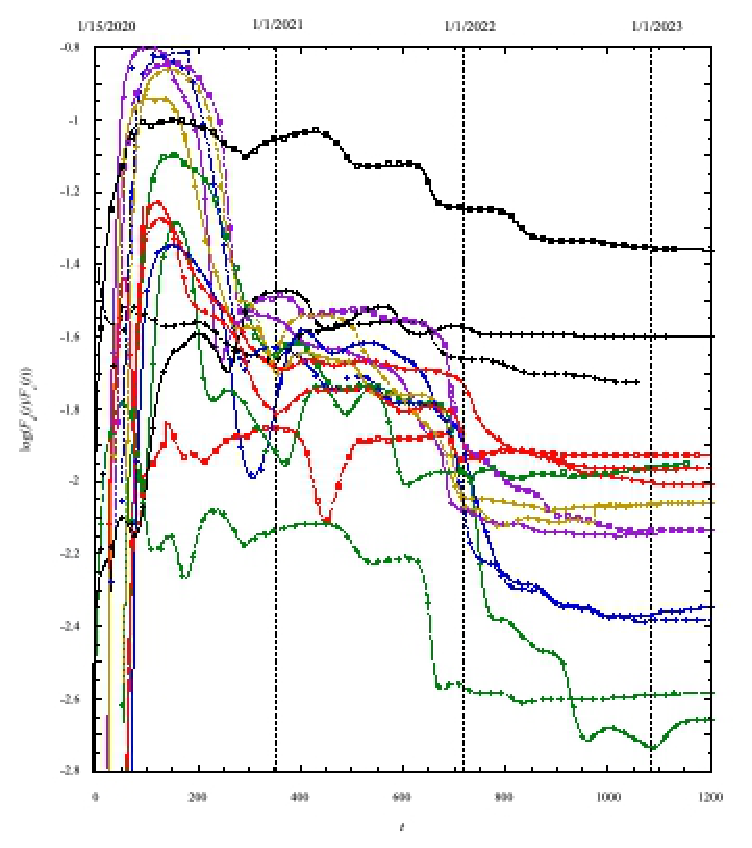}
\end{center}
\caption{(Color online) A log plot of the death rate $F_d(t)/F_c(t)$ versus time $t$. The symbols indicate the observed data for $F_d(t)/F_c(t)$ in different countries and the various solid lines the ratio $f_d(t)/f_c(t)$. The vertical dashed lines indicate the calendar days. The details are the same as in Fig. \ref{jp24}.}
\label{jp27}
\end{figure}

\subsection{Analytic calculations of $\langle  x_{\alpha}(t)^2\rangle$}
Next, we also compare the analytic results with the observed data.
In Fig. \ref{jp16}, the analytic results for $\langle  x_c(t)^2\rangle$ and $\langle  \delta n_c(t)^2\rangle$ are plotted versus time together with the observed data for $x_c(t)^2$ and $\delta n_c(t)^2$, respectively. In Fig. \ref{jp17}, the analytic results for $\langle  x_d(t)^2\rangle$ and $\langle  \delta n_d(t)^2\rangle$ are also plotted versus time together with the observed data for $x_d(t)^2$ and $\delta n_d(t)^2$, respectively. Here we have set $x_{\alpha}(t_{\alpha})=0$ (or $\chi_{\alpha}=1$) for simplicity. 
The jagged process is then seen to repeat at each stage. Since the only one observed data are available here, we directly compare them with the analytic results. Their dynamical behaviors seem to be approximately similar to each other. Finally, we mention here that since the numerical results for $x_{\alpha}(t)$ are repeatedly obtained by generating the random numbers for $\xi_{\alpha}(t)$ under the fixed value of $\gamma_{\alpha}$, one can easily find their averages at each stage and show that they coincide with the analytic results given Eq. (\ref{nellx}).

\subsection{Analytic calculations of $\langle  y_{\alpha}(t)^2\rangle$}
The analytic results of  $\langle  \delta f_{\alpha}(t)^2\rangle$ and $\langle  y_{\alpha}(t)^2\rangle$ are given by Eqs. (\ref{fave2}) and (\ref{yave2}), respectively. Similarly to Figs. \ref{jp16} and \ref{jp17}, therefore, we also compare them with the observed data. As is seen in Figs. \ref{jp16} and \ref{jp17}, the dynamics of $\delta n_{\alpha}(t)^2$ is well recovered by the dynamics of $\langle  \delta n_{\alpha}(t)^2\rangle$ within error, even though we have set $\langle x_{\alpha}(t_{\alpha})^2\rangle$=0. On the other hand, we have to choose the initial value $\langle y_{\alpha}(t_{\alpha})^2\rangle$ appropriately. This situation is different from that in $\langle  \delta n_{\alpha}(t)^2\rangle$ because of the fact that $f_{\alpha}(t) \gg n_{\alpha}(t)$. In order to recover the dynamics of $\delta f_c(t)^2$ reasonably, we adjust the peak value $\langle y_{\alpha}(t_{\alpha \ell}^{(i+1)}-1)^2\rangle$ at $i$th stage in the $\ell$th wave so as to coincide with the observed data for $y_{\alpha}(t_{\alpha \ell}^{(i+1)}-1)^2$. In Fig. \ref{jp18}, the analytic results for $\langle  y_c(t)^2\rangle$ and $\langle  \delta f_c(t)^2\rangle$ are then plotted versus time together with the observed data for $y_c(t)^2$ and $\delta f_c(t)^2$, respectively. In Fig. \ref{jp19}, the analytic results for $\langle  y_d(t)^2\rangle$ and $\langle  \delta f_d(t)^2\rangle$ are also plotted versus time together with the observed data for $y_d(t)^2$ and $\delta f_d(t)^2$, respectively. The dynamics of $\langle  \delta f_{\alpha}(t)^2\rangle$ is thus shown to recover that of $\delta f_{\alpha}(t)^2$ approximately well. Finally, we note that by using Eq. (\ref{dfs}), the numerical results for $y_{\alpha}(t)$ are also repeatedly obtained by generating the random numbers for $\xi_{\alpha}(t)$ under the fixed value of $\gamma_{\alpha}$ and using the initial value $y_{\alpha}(t_{\alpha})$ discussed above. Their averages are then shown to coincide with the analytic results such as Eq. (\ref{yave2}).

\subsection{Strong similarity between case and death}
Finally, we discuss the relation between the dynamics of case and that of death. As is shown in Figs. \ref{jp1} and \ref{jp2}, their causal motions are similar to each other. This suggests that the death counts are strongly related to the case counts in causal motions, just like a parents-child relationship. Next, we also investigate such a relation for their fluctuations. As discussed before, the fluctuations $\delta n_{\alpha}(t)$ and $\delta f_{\alpha}(t)$ are generated by the stochastic force $R_{\alpha}(t)$. In Fig. \ref{jp20}, we first compare the fluctuations obtained by the observed data for $\eta_d(t)$, $x_d(t)$, and $y_d(t)$ directly with those for $\eta_c(t)$, $x_c(t)$ and $y_c(t)$, respectively. Their dynamical behaviors are shown to be similar to each other, except the slight difference between their magnitudes. In Fig. \ref{jp21}, we then compare the fluctuations obtained by the observed data for $R_d(t)$, $\delta n_d(t)$, and $\delta f_d(t)$ directly with those for $R_c(t)$, $\delta n_c(t)$, and $\delta f_c(t)$, respectively. Their dynamical behaviors are also shown to be similar to each other, except their magnitudes which depend on $n_{\alpha}(t)$. We also compare the analytic results for case with those for death. In Fig. \ref{jp22}, we first plot the analytic results for $<x_{\alpha}(t)^2>$, $<y_{\alpha}(t)^2>$, and $d_{\alpha}(t)$. Their dynamical behaviors are shown to be similar to each other. In Fig. \ref{jp23}, we then plot the analytic results for $<\delta n_{\alpha}(t)^2>$, $<\delta f_{\alpha}(t)^2>$, and $D_{\alpha}(t)$. Their dynamical behaviors are again shown to be similar to each other, except their magnitudes which depend on $n_{\alpha}(t)$. Here we note that such a similarity is also seen for the other average quantities. Thus, those results suggest that there exists a strong similarity between case and death not only in causal motions but also in fluctuations. 

\section{Summary}
In the present paper, we have studied the non-equilibrium fluctuations in a number of COVID-19 cases and deaths observed in different communities from a statistical-dynamical point view. First, we have shown that they can be described by the stochastic equation (\ref{Meq}) whose stochastic force is a multiplicative type. In order to clarify the stochastic properties of the fluctuations, we have transformed it into the Langevin-type equation (\ref{Leq}) with the additive-type stochastic force $R_{\alpha}(t)$ together with the corresponding Fokker- Planck-type equation (\ref{FPeq}) by employing the time-convolutionless projection-operator method \cite{toku75,toku76}. Similarly, we have also derived the Fokker-Planck-type equation (\ref{FPfeq}). The average quantities $\langle  x_{\alpha}(t)^{\nu}\rangle$ and $\langle  y_{\alpha}(t)^{\nu}\rangle$ have been found analytically. Then, we have investigated the observed data not only analytically but also numerically from a unified point view. Thus, we have concluded that the present theory can describe not only the causal motion but also the dynamics of the non-equilibrium fluctuations. In this paper, we have shown only the theoretical analyses of the data observed in Japan on account of the space. However, we should mention that the present theoretical analyses are also applicable for the data observed in different communities. In fact, the unknown parameters $\rho_{\alpha}$ (or $\lambda_{\alpha}$), $t_{\alpha}$, $n_{\alpha}(t_{\alpha})$, and $f_{\alpha}(t_{\alpha})$ are determined consistently by fitting the analytic results for $f_{\alpha}(t)$ with observed data in each community. In Fig. \ref{jp24}, the fitting values of the infection rate $\rho_{\alpha}$ in different countries are plotted versus time. Thus, it turns out that the values of $\rho_{\alpha}$ exist between 0.90 and 1.30, except errors at earlier times. As seen in Fig. \ref{jp10}, the values of $Q_{\alpha}$ are also obtained in different communities. Once the fitting values of $t_{\alpha \ell}^{(i)}$ and $t_{\alpha \ell}^{(i+1)}$ are found, therefore, one can obtain the value of $\gamma_{\alpha \ell}^{(i)}$ from Eq. (\ref{ga2}). In Fig. \ref{jp25}, the values of $\gamma_{\alpha}$ in different countries are then plotted versus time. Hence the data observed in different communities are also expected to show similar behaviors to those discussed in Japan. In order to see one of such results, in Fig. \ref{jp26}, we plot the fluctuating cumulative numbers $F_{\alpha}(t)$ observed in various countries versus time together with the analytic results for $f_{\alpha}(t)$. Then, the observed data are shown to be well described by $f_{\alpha}(t)$ in any countries because of $|y_{\alpha}(t)|\ll 1$. Here we should note that although we have not discussed the fluctuations $\delta n_{\alpha}(t)$ and $\delta f_{\alpha}(t)$ observed in different countries on account of the space, their dynamics and their stochastic properties are also shown to be very similar to those in Japan. Hence a strong correlation between case and death for the data observed in different communities is easily expected to exist not only in causal motions but also in fluctuations. 

In Fig. \ref{jp27}, we next plot the death rates $F_d(t)/F_c(t)$ observed in different countries versus time together with the analytic results for $f_d(t)/f_c(t)$. They are shown to be well described by the ratios $f_d(t)/f_c(t)$ (see also Fig. \ref{jp7} (c)).  It turns out from Fig. \ref{jp27} that the death rates clearly decrease twice around April, 2020 and also around December, 2021. Such decays may be related to the COVID-19 vaccinations. Especially in Japan, a marked decrease is seen around January 2022. In fact, according to the government, the total number of COVID-19 vaccinations in Japan exceeded 300 million as of October 2022.

Finally, we would like to offer our condolences to the many innocent people around the world who have lost their lives due to this coronavirus pandemic, and we would also like express our heartfelt gratitude to the many medical workers and others involved in the relief of coronavirus patients. We sincerely hope that the present research will be of some help in preventing the next pandemic.

\appendix
\section{Derivation of Eq. (\ref{Ome})}
In the present system, there exist three types of peoples. One is an infected person (or a dead person) whose total number is given by $N_{\alpha}(t)$ at time $t$. The second is a close contact person with infected persons, whose total number is given by $C_{\alpha}(t)$. The last is an uninfected person whose total number is given by $P(t)-N_{\alpha}(t)-C_{\alpha}(t)$, where $P(t)$ is a population of the target area. Since the infected (or dead) persons further increase by the close contact persons, we assume that $N_{\alpha}(t)$ is described by the following equation:
\begin{equation}
\frac{d}{dt}N_{\alpha}(t)=\lambda_{\alpha 0} N_{\alpha}(t)+\kappa_{\alpha}C_{\alpha}(t)N_{\alpha}(t), \label{Meq0}
\end{equation}
where $\kappa_{\alpha}$ is a positive constant. Since the close contact persons are drifting around the uninfected persons, the relation between them seems to be similar to that between the Brownian particle and the bath. Hence one may assume that $C_{\alpha}(t)$ obeys a Langevin-type equation given by
\begin{equation}
\frac{d}{dt}C_{\alpha}(t)=-\zeta_{\alpha} C_{\alpha}(t)+\omega_{\alpha}(t), \label{Meq1}
\end{equation}
where $\zeta_{\alpha}$ is a positive constant whose magnitude is assumed to be much larger than that of  $\lambda_{\alpha 0}$. Here $\omega_{\alpha}(t)$ is a random force of an additive type and is assumed to satisfy
\begin{equation}
\langle\omega_{\alpha}(t)\rangle=\langle\omega_{\alpha}(t)C_{\alpha}(t_{\alpha})\rangle=0.\label{Meq2}
\end{equation}
Here $\omega_{\alpha}(t)$ is also assumed to satisfy the fluctuation-dissipation relation given by
\begin{equation}
\langle\omega_{\alpha}(t)\omega_{\alpha}(t')\rangle=2\zeta_{\alpha}\langle C_{\alpha}(t_{\alpha})^2\rangle\delta(t-t'). \label{Meq3}
\end{equation}
From Eq. (\ref{Meq1}), one then obtains 
\begin{eqnarray}
\langle C_{\alpha}(t)\rangle&=&e^{-\zeta_{\alpha}(t-t_{\alpha})}\langle C_{\alpha}(t_{\alpha})\rangle, \label{C1}\\
\langle C_{\alpha}(t)^2\rangle&=&\langle C_{\alpha}(t_{\alpha})^2\rangle. \label{C2}
\end{eqnarray}
On the time scale larger than $\tau_c(=\zeta_{\alpha}^{-1})$, therefore, $\langle C_{\alpha}(t)\rangle$ goes to zero. Since $\lambda_{\alpha 0}\ll \zeta_{\alpha}$, on the time scale of $\tau_{\alpha 0}(=\lambda_{\alpha 0}^{-1})$, one can then solve Eq. (\ref{Meq1}) to find
\begin{equation}
C_{\alpha}(t)\simeq \omega_{\alpha}(t)/\zeta_{\alpha}. \label{Meq4}
\end{equation}
From Eqs. (\ref{Meq}) and (\ref{Meq0}), one can then write the stochastic force $\Omega_{\alpha}(t)$ as
\begin{equation}
\Omega_{\alpha}(t) =\xi_{\alpha}(t)N_{\alpha}(t) \label{Meq5}
\end{equation}
with the random force $\xi_{\alpha}(t)=(\kappa_{\alpha}/\zeta_{\alpha})\omega_{\alpha}(t)$. From Eqs. (\ref{xi}) and (\ref{Meq3}), one can also find $\gamma_{\alpha}=\kappa_{\alpha}^2/\zeta_{\alpha}$.


\begin{references}
\bibitem{KM27}W. O. Kermack and A. G. McKendrick, Contributions to the mathematical theory of epidemics I, Proceedings of the Royal Society {\bf115A} (1927) 700.
\bibitem{KM32}W. O. Kermack and A. G. McKendrick, Contributions to the mathematical theory of epidemics II. The problem of endemicity, Proceedings of the Royal Society {\bf138A} (1932) 55.
\bibitem{DHB13}O. Diekmann, J. A. P. Heesterbeek, and T. Britton (2013), Mathematical Tools for Understanding Infectious Disease Dynamics, Princeton University Press, Princeton and Oxford.
\bibitem{toku75} M. Tokuyama and H. Mori, Prog. Theor. Phys. {\bf54} (1975) 918.
\bibitem{toku76} M. Tokuyama and H. Mori, Prog. Theor. Phys. {\bf55} (1976) 411.
\bibitem{kubo63} R. Kubo, J. Math. Phys. {\bf4} (1963) 174.
\bibitem{toku80} M. Tokuyama, Physica A {\bf102} (1980) 399.
\bibitem{toku81} M. Tokuyama, Physica A {\bf109} (1981) 128.


\end{references}
\end{document}